\documentstyle[11pt,amssymb]{article}
\begin{document}


\newcommand{\nnl}{\nl[6mm]}
\newcommand{\enl}{\\[6mm]}
\newcommand{\nllbnl}[1]{\nl[-3mm]\label{#1}\\[-3mm]}
\newcommand{\nle}{\nl[-3mm]\\[-3mm]}
\newcommand{\nl}{\nonumber\\}
\newcommand{\bl}{&&\quad}
\newcommand{\ab}{\allowbreak}

\newcommand{\iint}{{\int\hskip-3mm\int}}
\newcommand{\iiint}{{\int\hskip-3mm\int\hskip-3mm\int}}
\newcommand{\dddot}[1]{{\mathop{#1}\limits^{\vbox to 0pt{\kern 0pt
 \hbox{.{\kern-0.25mm}.{\kern-0.25mm}.}\vss}}}}
\renewcommand{\leq}{\leqslant}
\renewcommand{\geq}{\geqslant}

\renewcommand{\theequation}{\thesection.\arabic{equation}}
\let\ssection=\section
\renewcommand{\section}{\setcounter{equation}{0}\ssection}

\newcommand{\be}{\begin{equation}}
\newcommand{\ee}{\end{equation}}
\newcommand{\bes}{\begin{eqnarray}}
\newcommand{\ees}{\end{eqnarray}}
\newcommand{\eens}{\nonumber\end{eqnarray}}

\renewcommand{\/}{\over}
\renewcommand{\d}{\partial}
\newcommand{\dd}{{\check\d}}
\newcommand{\ddt}[1]{{\d{#1}\/\d t}}
\newcommand{\dNx}{d^N\!x\ }
\newcommand{\dNy}{d^N\!y\ }
\newcommand{\dNz}{d^N\!z\ }

\newcommand{\eps}{\epsilon}
\newcommand{\gm}{\gamma}
\newcommand{\dlt}{\delta}
\newcommand{\th}{\theta}
\newcommand{\al}{\alpha}
\newcommand{\bt}{\beta}
\newcommand{\si}{\sigma}
\newcommand{\ka}{\kappa}
\newcommand{\la}{\lambda}
\newcommand{\ups}{\upsilon}

\newcommand{\tdiff}{{\widetilde{diff}}}

\newcommand{\sd}{\,{\rm sd}}
\newcommand{\tr}{{\rm tr}}
\newcommand{\str}{{\rm str}}
\newcommand{\diff}{{\rm diff}}
\newcommand{\rrep}{{\rm rep}}
\newcommand{\gauge}{{\rm gauge}}

\newcommand{\gloj}{gl(N)\oplus\oj}
\newcommand{\sqg}{\sqrt{|g(x)|}\,}

\newcommand{\xz}{\xi^0}
\newcommand{\yz}{\eta^0}
\newcommand{\xx}{{\mathbf x}}
\newcommand{\bb}{{\frak b}}
\newcommand{\cc}{{\frak c}}

\newcommand{\xmu}{\xi^\mu}
\newcommand{\xnu}{\xi^\nu}
\newcommand{\xrho}{\xi^\rho}
\newcommand{\xsi}{\xi^\si}
\newcommand{\ynu}{\eta^\nu}
\newcommand{\dmu}{\d_\mu}
\newcommand{\dnu}{\d_\nu}
\newcommand{\dsi}{\d_\si}
\newcommand{\dtau}{\d_\tau}
\newcommand{\drho}{\d_\rho}
\newcommand{\qmu}{q^\mu}
\newcommand{\qnu}{q^\nu}
\newcommand{\qrho}{q^\rho} 
\newcommand{\qsi}{q^\si}
\newcommand{\qtau}{q^\tau}
\newcommand{\pmu}{p_\mu}
\newcommand{\pnu}{p_\nu}

\newcommand{\hL}{{\hat L}}
\newcommand{\hphi}{{\hat\phi}}
\newcommand{\hpi}{{\hat\pi}}

\newcommand{\tpi}{{1\/2\pi i}}
\newcommand{\half}{{1\/2}}
\newcommand{\Np}[1]{{N+p\choose p #1}}

\newcommand{\mm}{{\mathbf m}}
\newcommand{\nn}{{\mathbf n}}
\newcommand{\rr}{{\mathbf r}}
\newcommand{\mmu}{{\underline \mu}}
\newcommand{\nnu}{{\underline \nu}}
\newcommand{\noll}{{\underline 0}}
\newcommand{\one}{{\underline 1}}

\newcommand{\fa}{\phi_\alpha}
\newcommand{\fb}{\phi_\beta}
\newcommand{\pa}{\pi^\alpha}
\newcommand{\pb}{\pi^\beta}
\newcommand{\fsa}{\phi^{*\alpha}}
\newcommand{\fsb}{\phi^{*\beta}}
\newcommand{\psa}{\pi^*_\alpha}
\newcommand{\psb}{\pi^*_\beta}
\newcommand{\Ea}{\EE^\alpha}
\newcommand{\Eb}{\EE^\beta}
\newcommand{\Eam}{\Ea_{,\mm}}

\newcommand{\fm}[1]{\phi_{#1,\mm}}
\newcommand{\fn}[1]{\phi_{#1,\nn}}
\newcommand{\pim}[1]{\pi^{#1,\mm}}
\newcommand{\pin}[1]{\pi^{#1,\nn}}
\newcommand{\fsm}[1]{\phi^{*#1}_{,\mm}}
\newcommand{\fsn}[1]{\phi^{*#1}_{,\nn}}
\newcommand{\psm}[1]{\pi^{*,\mm}_{#1}}
\newcommand{\psn}[1]{\pi^{*,\nn}_{#1}}
\newcommand{\bsm}[1]{\bt^{*#1}_{,\mm}}
\newcommand{\csm}[1]{\gm^{*,\mm}_{#1}}

\newcommand{\summ}[1]{{\sum_{|\mm|\leq #1}}}
\newcommand{\sumn}[1]{{\sum_{|\nn|\leq #1}}}
\newcommand{\summn}[1]{{\sum_{|\mm|\leq|\nn|\leq #1}}}
\newcommand{\summrn}{{\sum_{|\mm|\leq|\rr|\leq|\nn|}}}

\newcommand{\map}[1]{{\ \stackrel{#1}{\longrightarrow}\ }}
\newcommand{\Hstate}{H^\bullet_{state}}

\newcommand{\LL}{{\cal L}}
\newcommand{\J}{{\cal J}}
\newcommand{\MM}{{\cal M}}
\newcommand{\DD}{{\cal D}}
\newcommand{\EE}{{\cal E}}
\newcommand{\FF}{{\cal F}}
\newcommand{\GG}{{\cal G}}
\newcommand{\PP}{{\cal P}}
\newcommand{\QQ}{{\cal Q}}
\newcommand{\KK}{{\cal KT}}
\newcommand{\NS}{{\cal N}_S}
\newcommand{\BB}{{\cal B}}
\newcommand{\OO}{{\cal O}}
\newcommand{\UU}{{\cal U}}
\newcommand{\SS}{{\cal S}}
\newcommand{\TT}{{\cal T}}
\newcommand{\WW}{{\cal W}}

\newcommand{\CF}{C^\infty}
\newcommand{\QKT}{Q_{KT}}
\newcommand{\Omb}{\Omega^\bullet_{KT}}
\newcommand{\dimm}{\dim\kern0.1mm}

\newcommand{\Lxi}{\LL_\xi}
\newcommand{\Leta}{\LL_\eta}
\newcommand{\gh}{{{\rm gh}}}
\newcommand{\mom}{{{\rm mom}}}
\newcommand{\tot}{{{\rm TOT}}}

\newcommand{\bosonic}{\hbox{{\scriptsize bosonic}}}
\newcommand{\fermionic}{\hbox{{\scriptsize fermionic}}}
\newcommand{\auxiliary}{\hbox{{\scriptsize auxiliary}}}

\newcommand{\cq}{c^{(q)}}
\newcommand{\ce}{c^{(e)}}
\newcommand{\cf}{c^{(\phi)}}
\newcommand{\tcf}{\tilde c^{(\phi)}}

\newcommand{\into}{\hookrightarrow}
\newcommand{\e}{{\rm e}}
\newcommand{\im}{{\rm im}\,}
\newcommand{\rep}{\varrho}
\newcommand{\ad}{{\rm ad}}
\newcommand{\Ad}{(1,0;0)\oplus\ad_\oj}

\newcommand{\oj}{{\frak g}}

\newcommand{\mum}{{\mu_1..\mu_m}}
\newcommand{\nun}{{\nu_1..\nu_n}}
\newcommand{\rnun}{{\rho|\nu_1..\nu_n}}
\newcommand{\nux}[1]{{\nu_1..#1..\nu_n}}
\newcommand{\nunj}{{\nu_1..\check\nu_j..\nu_n}}

\newcommand{\bra}[1]{\big{\langle}#1\big{|}}
\newcommand{\ket}[1]{\big{|}#1\big{\rangle}}
\newcommand{\phys}{{\rm phys}}
\newcommand{\Noether}{{\rm Noether}}

\newcommand{\km}[1]{{\widehat{#1}}}
\newcommand{\no}[1]{{\,:\kern-0.7mm #1\kern-1.2mm:\,}}

\newcommand{\RR}{{\Bbb R}}
\newcommand{\CC}{{\Bbb C}}
\newcommand{\ZZ}{{\Bbb Z}}

\topmargin 1.0cm

\newpage
\vspace*{-3cm}
\pagenumbering{arabic}
\begin{flushright}
{\tt math-ph/9908028}
\end{flushright}
\vspace{12mm}
\begin{center}
{\huge Quantum physics as the projective representation theory of 
Noether symmetries}\\[8mm]
\renewcommand{\baselinestretch}{1.2}
\renewcommand{\footnotesep}{10pt}
{\large T. A. Larsson\\
}
\vspace{5mm}
{\sl Vanadisv\"agen 29\\
S-113 23 Stockholm, Sweden}\\
email: tal@hdd.se\\
\end{center}
\vspace{3mm}
\begin{abstract}
I construct lowest-energy representations of non-centrally extended
algebras of Noether symmetries, including diffeomorphisms and 
repara\-metrizations of the observer's trajectory.
This may be viewed as a new scheme for quantization. 
First classical physics is formulated as the cohomology of a certain 
Koszul-Tate (KT) complex, using not only fields and antifields but also
their conjugate momenta. Then all fields are expanded in a Taylor series
around the observer's present position, and terms of order higher
than $p$ are truncated. Finally, quantization is carried out by
replacing Poisson brackets by commutators and imposing the KT
cohomology in Fock space. This procedure is consistent
for finite $p$, but the limit $p\to\infty$ leads to difficulties.
\end{abstract}

\vspace{5mm}

\section{Introduction}

The main hypothesis underlying this work is that physics is the
representation theory of its Noether symmetries, the most prominent
ones being spacetime diffeomorphisms and repara\-metrizations of the
observer's trajectory. After quantization one expects to find a
projective representation of this group, i.e. a representation up 
to a local phase. On the Lie algebra level, this corresponds
to an abelian but non-central extension of $diff(N)\oplus diff(1)$;
only if the phase is globally constant, the Lie algebra extension
is central.
In \cite{Lar98}, I discovered the ``DRO 
(Diffeomorphism, Repara\-metrization, Observer) algebra'' 
$DRO(N)$ (the name, however, is new), which is a
non-split abelian extension of $diff(N)\oplus diff(1)$
by the commutative algebra of local functionals on the observer's 
trajectory, depending on four parameters (``abelian charges'').
This discovery builds on previous work by Eswara-Rao and 
Moody \cite{ERM94} and myself \cite{Lar97}.
Related work goes under the name ``toroidal Lie algebras''
\cite{BB98, Bil97, EMY92, MEY90}. 
The DRO algebra, and the more general ``DGRO (Diffeomorphism, Gauge,
Repara\-metrization, Observer) algebra'' $DGRO(N, \oj)$ obtained by
adding an ordinary gauge algebra $map(N,\oj)$ to the Noether symmetries,
are described in section 2.

The main ingredient missing in \cite{Lar98} is that it makes no
reference to dynamics (i.e. action, Hamiltonian, or Euler-Lagrange (EL)
equations), so it is a purely kinematical theory. To introduce
dynamics into the picture, I employ the following strategy.
First the solutions to the classical equations of motion 
(EL and geodesic) are described in terms of the cohomology of 
a certain Koszul-Tate (KT) complex. This is closely related to
the Batalin-Vilkovisky formalism, as formulated in
\cite{HT92, Sta97}. However, there is one important
difference: I do not only introduce fields and antifields, but also
field and antifield momenta. This has several implications: 
1. The KT differential can be expressed as a Poisson bracket with a
KT generator.
2. The antibracket is not a fundamental object.
3. The cohomology grows; it consists of differential forms 
(not just functions) on the stationary surface. 
Nevertheless, this enlarged KT cohomology still
encodes classical dynamics, and it is the subject of section 3.

The next idea is to expand all fields and antifields, but
not their momenta, in a Taylor series around the observer's present 
position, and to truncate after terms of order $p$,
i.e. to pass to $p$-jet space. 
I now quantize in the na\"\i ve sense of the word:
take a formulation of classical physics, replace all Poisson brackets
by commutators and representent the resulting Heisenberg algebra on
a unique Fock space.
Although the fields do not depend on ``parameter time'' (i.e. the
parameter along the observer's trajectory), the Taylor coefficients
do. It is therefore possible to make a Fourier expansion of both 
the jets, the trajectory, and all momenta in
parameter time, and proclaim that the Fock vacuum be annihilated by
all negative energy modes. This step, and in particular the
form of the resulting extensions, was the main result of \cite{Lar98};
it is reviewed in section 4.

The main advantage of the KT complex is that it survives quantization.
In contradistinction to the BRST generator, the KT generator is bilinear in 
commuting variables, and thus already normal ordered. In section 5
I describe the KT generator in jet space and the associated quantum
KT complex. In particular, the action of the DGRO algebra on the
cohomology is computed. 
The resulting $DGRO(N,\oj)$ modules are manifestly 
well-defined quantum theories for all finite $p$. This can be viewed
as an extreme statement of locality: the theory only deals with
objects that are local to the observer, i.e. the fields and finitely
many derivatives thereof at the observer's present position.
This is not to say that events away from the observer are unphysical,
but they are not described by the theory. To recover
objective reality of distant events, we should demand that the limit
$p\to\infty$ exists. The leading behaviour of the abelian charges
is studied, but it is found to diverge due to second-order antifields.
Some means to avoid this type of infinity are discussed, but none of 
these is satisfactory.

In the course of this work I introduce several modifications to the
formalism of physics. These changes are dictated by the desire to
obtain well-defined quantum representations of the Noether symmetries,
but there remains to clarify the relation to standard formulations
of quantum physics. However, even if my results turn out to be 
physically irrelevant, they are still of independent mathematical 
interest since new representations of naturally arising Lie algebras
are constructed.

\section{ The algebras $DRO(N)$ and $DGRO(N,\oj)$ }
\label{sec:DRO}

Let $\xi=\xmu(x)\dmu$, $x\in\RR^N$, $\dmu = \d/\d x^\mu$,
be a vector field, with commutator 
$[\xi,\eta] \equiv \xmu\dmu\ynu\dnu - \ynu\dnu\xmu\dmu$.
Greek indices $\mu,\nu = 0,1,..,N\!-\!1$ label the 
spacetime coordinates and the summation convention is used on all kinds 
of indices.
The diffeomorphism algebra (algebra of vector fields, Witt algebra) 
$diff(N)$ is generated by Lie derivatives $\Lxi$.
In particular, we refer to diffeomorphisms on the circle as 
repara\-metrizations. They form an additional $diff(1)$ algebra with 
generators $L_f$, where $f = f(t)d/dt$, $t\in S^1$, is a vector field
on the circle. The commutator is $[f,g] = (f\dot g - g\dot f)d/dt$,
where a dot indicates the $t$ derivative.
Moreover, introduce $N$ priviledged functions on the circle $\qmu(t)$,
which can be interpreted as the trajectory of an observer (or base point).
Let the observer algebra $Obs(N) = \CC[q(t)]$ be
the space of local functionals of $\qmu(t)$, i.e. polynomial functions of  
$\qmu(t)$, $\dot\qmu(t)$, ... $d^k \qmu(t)/dt^k$,  $k$ finite, 
regarded as a commutative Lie algebra.

The assumption that $t\in S^1$ is for technical simplicity; it enables
jets to be expanded in a Fourier series, but it is physically
quite unjustified because it means that spacetime is periodic in
the time direction. However, all we really need is that
$\int dt\ \dot F(t) = 0$ for all functions $F(t)$. Most results are
unchanged if we instead take $t\in\RR$ and replace Fourier sums with
Fourier integrals everywhere.

The {\em DRO (Diffeomorphism, Repara\-metrization, Observer)} algebra 
\break 
$DRO(N)$ is an abelian but non-central Lie algebra extension of
$diff(N)\oplus diff(1)$ by $Obs(N)$:
\be
0 \map{} Obs(N) \map{} DRO(N) \map{}
 diff(N)\oplus diff(1) \map{} 0.
\label{seq}
\ee
The extension depends on the four parameters $c_j$, $j = 1, 2, 3, 4$,
to be called  {\em abelian charges}; the name is chosen in analogy
with the central charge of the Virasoro algebra.
The sequence (\ref{seq}) splits ($DRO(N)$ is a semi-direct product)
iff all four abelian charges vanish. The brackets are given by
\bes
[\Lxi,\Leta] &=& \LL_{[\xi,\eta]} 
 + \tpi\int dt\ \dot\qrho(t) 
 \Big( c_1 \drho\dnu\xmu(q(t))\dmu\ynu(q(t)) +\nl
 \bl+ c_2 \drho\dmu\xmu(q(t))\dnu\ynu(q(t)) \Big), \nl
{[}L_f, \Lxi] &=& {c_3\/4\pi i} \int dt\ 
 (\ddot f(t) - i\dot f(t))\dmu\xmu(q(t)), \nl
{[}L_f,L_g] &=& L_{[f,g]} 
 + {c_4\/24\pi i}\int dt (\ddot f(t) \dot g(t) - \dot f(t) g(t)), 
\label{DRO}\\
{[}\Lxi,\qmu(t)] &=& \xmu(q(t)), \nl
{[}L_f,\qmu(t)] &=& -f(t)\dot\qmu(t), \nl
{[}\qmu(s), \qnu(t)] &=& 0,
\eens
extended to all of $Obs(N)$ by Leibniz' rule and linearity.
Two abelian charges have been renamed compared to \cite{Lar98}:
$c_3 = c_0$ and $c_4 = c$, where $c$ is the central 
charge in the Virasoro algebra generated by repara\-metrizations.
Also, the value of a trivial cocycle has been fixed.

To prove that (\ref{DRO}) defines a Lie algebra is straightforward;
one either checks all Jacobi identities, or notes the existence of the
explicit realization below. Non-triviality was not
proven in \cite{Lar98}, but this is easily rectified. The strategy is
to consider the restriction of (\ref{DRO}) to various subalgebras.
The $L_f$ generate a Virasoro algebra with central charge $c_4$,
and the terms proportional to $c_1$ and $c_2$
are identified as extensions $\psi^W_4$ and $\psi^W_3$ in 
Dzhumadil'daev's classification \cite{Dzhu96}. To prove that $c_3$
is non-trivial, we set $c_1=c_2=c_4=0$ and consider the restriction to
the subalgebra generated by $K_f = \Lxi+L_f$, where $\xi = f(x^0)\d_0$:
\bes
[K_f,K_g] &=& K_{[f,g]} + {c_3\/4\pi i} \int dt\ 
(\ddot f(t) g'(q^0(t)) - f'(q^0(t))\ddot g(t)), 
\label{KK}\\
{[}K_f, q^0(t)] &=& f(q^0(t)) - \dot q^0(t) f(t),
\eens
apart from a trivial term. If we (consistently) set $q^0(t)=t$,
(\ref{KK}) becomes a Virasoro algebra with central charge $12c_3$,
and hence $c_3$ is non-trivial.
Finally, we note that all four terms behave differently
under the restrictions considered, so they must be inequivalent. Q.E.D.

It is not difficult to reformulate the DRO algebra as a proper Lie algebra,
by introducing a compete basis for $Obs(N)$. In fact, it suffices to
consider two infinite families of linear operators 
$S_n^\nun(F_\nun)$, $R_n^\rnun(G_\rnun)$, defined
for arbitrary functions $F_\nun(t,x)$, $G_\rnun(t,x)$, $t\in S^1$, 
$x\in\RR^N$, totally symmetric in the indices $\nun$.
\bes
S_n^\nun(F_\nun) &=& \tpi \int dt\ \dot q^{\nu_1}(t) .. \dot q^{\nu_n}(t) 
 F_\nun(t,q(t)), 
\nllbnl{SR}
R_n^\rnun(G_\rnun) &=& \tpi \int dt\ \ddot q^\rho(t) 
 \dot q^{\nu_1}(t) .. \dot q^{\nu_n}(t) G_\rnun(t,q(t)).
\eens
Then $\Lxi$, $L_f$, $S_n^\nun(F_\nun)$,
$R_n^\rnun(G_\rnun)$ generate a Lie algebra, whose brackets are 
explicitly written down in \cite{Lar98}. 

Consider also the gauge (or current) algebra $map(N,\oj)$,
where $\oj$ is finite-dimen\-sional Lie algebra with basis $J^a$ 
(hermitian if $\oj$ is compact and semisimple), 
structure constants $f^{ab}{}_c$ and brackets
$[J^a,J^b] = if^{ab}{}_c J^c$. 
Our notation is similar to \cite{GO86} or \cite{FMS96}, chapter 13.
We always assume that $\oj$ has a
Killing metric proportional to $\dlt^{ab}$. Then there is no need to
distinguish between upper and lower $\oj$ indices, and the structure
constants $f^{abc} = \dlt^{cd} f^{ab}{}_d$ are totally antisymmetric.
Further assume that there is a priviledged vector 
$\dlt^a \propto \tr J^a$, such that $f^{ab}{}_c \dlt^c \equiv 0$.
Of course, $\dlt^a = 0$ if $\oj$ is semisimple, but it may be non-zero
if $\oj$ contains abelian factors. The primary example is $\oj = gl(N)$,
where $\tr(T^\mu_\nu) \propto \dlt^\mu_\nu$.

Let $X=X_a(x)J^a$, $x\in\RR^N$, be a $\oj$-valued
function and define $[X,Y] = if^{ab}{}_c X_aY_bJ^c$.
$map(N,\oj)$ is the algebra of maps from $\RR^N$ to 
$\oj$. Its generators are denoted by $\J_X$.
The {\em DGRO (Diffeomorphism, Gauge, Repara\-metrization, Observer) 
algebra} $DGRO(N,\oj)$ has brackets
\bes
[\J_X, \J_Y] &=& \J_{[X,Y]} 
 - {c_5\/2\pi i} \dlt^{ab}\int dt\ \dot\qrho(t)\drho X_a(q(t))Y_b(q(t)), \nl
{[}L_f,\J_X] &=& {c_6 \/4\pi i}\dlt^a 
 \int dt\ (\ddot f(t) - i \dot f(t)) X_a(q(t)),
\label{DGRO} \\
{[}\Lxi, \J_X] &=& \J_{\xmu\dmu X} 
- {c_7 \/2\pi i} \dlt^a\int dt\ \dot\qrho(t) X_a(q(t))\drho\dmu\xmu(q(t)), \nl
{[}\J_X, \qmu(t)] &=& 0, 
\eens
in addition to (\ref{DRO}). In \cite{Lar98}, the constants were 
denoted by $k = c_5$ (the extended algebra reduces to the Kac-Moody
algebra $\km\oj$ when $N=1$ and $q^0(t) = t$), 
$g^a = c_6\dlt^a$ and ${g'}^a = c_7\dlt^a$.
The present notation has the advantage that all abelian charges
$c_j$, $j = 1, \ldots, 7$, can be discussed collectively.

It is sometimes better not to work with smeared generators, so we 
define $\LL_\mu(x)$, $L(t)$ and $\J^a(x)$ by
\bes
\Lxi &=& \int \dNx \xmu(x)\LL_\mu(x), \nl
L_f &=& \int dt\ f(t)L(t), 
\label{local}\\
\J_X &=& \int \dNx X_a(x) \J^a(x).
\eens

In \cite{Lar97} I described a gauge-fixed version of the DRO algebra,
denoted by $\tdiff(N)$. To obtain it, we must recall Dirac's treatment
of constrained Hamiltonian systems, and adapt it to Lie algebras
\cite{HT92}.
Consider embeddings of some Lie algebra $\oj$ into the
Poisson algebra $\CF(\PP)$, where $\PP$ is a phase space.
Let $P,R,..$ label constraints $\chi_P$, which are assumed bosonic
for simplicity.
Consider the constraint surface $\chi_P\approx0$,
where weak equality (i.e. equality
modulo constraints) is denoted by $\approx$.
Constraints are second class if the Poisson bracket matrix 
$C_{PR} = [\chi_P,\chi_R]$ 
is invertible; otherwise, they are first class and generate a Lie algebra.
Assume that all constraints are second class. Then the matrix 
$C_{PR}$ has an inverse, denoted by $\Delta^{PR}$.
The Dirac bracket
\be
[A,B]^* = [A,B] - [A,\chi_P] \Delta^{PR} [\chi_R,B]
\label{Dirac}
\ee
defines a new Poisson bracket which is compatible with the 
constraints: $[A, \chi_R]^* = 0$ for every $A\in \oj$.
Of course, there is no guarantee that the operators $A,B$ still generate
the same Lie algebra under the Dirac brackets. A sufficient condition
for this is that the constraints are preserved in the sense that
$[A,\chi_P] \approx 0$ for every $A$. A less restrictive condition is 
often possible. Usually, the constraints can be divided into two sets
$\chi_P = (\Phi_a, \Pi^a)$, such that 
$[\Phi^a, \Phi_b] \approx 0$.
The $\Phi^a$ are then first class, and  $\Pi_a$ are gauge conditions.
It is then sufficient that $[A, \Phi_a] \approx 0$, 
because the components of $\Delta^{PR}$ that involve $\Pi$'s on 
both sides vanish.  

Now consider the case $\oj=DRO(N)$. Strictly speaking, we can only pass
to Dirac brackets if $\oj$ admits a Poisson bracket realization, which
is not necessarily true in the presence of abelian extensions.
If we ignore this problem, repara\-metrizations and one component of 
the observer's trajectory can be eliminated 
by introduction of the second-class constraints
\be
\chi(s) = \pmatrix{ q^0(s) \cr L(s) } \approx 0.
\label{gcond}
\ee
In the absense of extensions, $[\Lxi,L(t)]=0$, so this constraint is of
the type above. When the extensions are turned on, new terms arise, 
but we still have an abelian extension of $diff(N)$.
The Poisson bracket matrix $C(s,t)$ and its inverse $\Delta(s,t)$ are,
on the constraint surface,
\bes
C(s,t) &\equiv& [\chi(s),\chi^T(t)]
= \Big[\pmatrix{q^0(s) \cr L(s) }, \pmatrix{q^0(t) & L(t) }\Big] \nl
&\approx& \pmatrix{ 0 & \dlt(s-t) \cr -\dlt(s-t) &
 {c_4\/24\pi i}(\dddot\dlt(s-t) + \dot\dlt(s-t)) }, \\
\Delta(s,t) &\approx& \pmatrix{ 
 {c_4\/24\pi i}(\dddot\dlt(s-t) + \dot\dlt(s-t)) & -\dlt(s-t) \cr
 \dlt(s-t) & 0 }.
\eens
We find
\bes
[\Lxi,\Leta]^* &=& \LL_{[\xi,\eta]} 
 + \tpi\int dt\  
  c_1 \dnu\dot\xmu(q(t))\dmu\ynu(q(t)) +\nl
 \bl+ c_2 \dmu\dot\xmu(q(t))\dnu\ynu(q(t)) +\nl
&&+ {c_3\/4\pi i} \int dt\ 
 \dnu\ynu(q(t)) ( \ddot\xz(q(t)) - i \dot\xz(q(t))) -\nl
\bl-\dmu\xmu(q(t)) (\ddot\yz(q(t)) - i\dot\yz(q(t)) ) +
\label{tdiff}\\
&&+ {c_4\/24\pi i} \int dt\ \ddot\xz(q(t))\dot\yz(q(t))
 - \dot\xz(q(t))\yz(q(t)), \nl
{[}\Lxi,\qmu(t)]^* &=& \xmu(q(t)) - \dot\qmu(t)\xz(q(t)), \nl
{[}\qmu(s), \qnu(t)]^* &=& 0, \nl
{[}L(s), \Lxi]^* &=& [L(s),L(t)]^* = [L(s),\qmu(t)]^* = 0.
\eens
Note that $[\Lxi, q^0(t)]^* = 0$. 
This algebra deserves to be called the gauge-fixed diffeomorphism 
algebra, denoted by $\tdiff(N)$.
Again, (\ref{tdiff}) is not a Lie algebra, but can be made so by
introducing the new generators (\ref{SR}) \cite{Lar97}.

\section{ Classical physics as Koszul-Tate cohomology }
\label{sec:class}

\subsection{ Configuration space and phase space }
The configuration space $\QQ$ is the space spanned by the
observer's trajectory $\qmu(t)$, $t\in S^1$, the einbein
$e(t)$, $t\in S^1$, and a collection of $V$-valued fields 
over spacetime, where $V$ carries a finite-dimen\-sional $gl(N)$ 
representation $\rep$.
The fields are collectively denoted by $\fa(x)$, $x\in\RR^N$, 
where the $V$ index $\al$ labels different tensor and internal 
components.
If $V$ contains several (bosonic or fermionic) field species,
$\al$ labels these as well; in this case 
$\rep = \rep_1 \oplus \ldots \oplus \rep_n$ is a direct sum.

In our convention, $gl(N)$ has basis $T^\mu_\nu$ and brackets
\be
[T^\mu_\nu,T^\si_\tau] = 
\dlt^\si_\nu T^\mu_\tau - \dlt^\mu_\tau T^\si_\mu.
\label{glN}
\ee
To these elements correspond the matrices $\rep(T^\nu_\mu)$ 
with elements $\rep^\al_\bt(T^\nu_\mu)$.
In particular, denote by $\rep = (p,q;\ka)$ the representation on
tensor densities with $p$ upper 
and $q$ lower indices and weight $\ka$:
\be
\rep(T^\mu_\nu)\phi^{\si_1..\si_p}_{\tau_1..\tau_q}
= -\ka \dlt^\mu_\nu \phi^{\si_1..\si_p}_{\tau_1..\tau_q} 
+\sum_{i=1}^p \dlt^{\si_i}_\nu 
 \phi^{\si_1..\mu..\si_p}_{\tau_1..\tau_q}
-\sum_{j=1}^q \dlt^\mu_{\tau_j}
 \phi^{\si_1..\si_p}_{\tau_1..\nu..\tau_q}.
\label{gltens}
\ee

{F}rom $\QQ$ we construct the corresponding phase space $\PP$
by adjoining the conjugate momenta $\pmu(t)$, $\pi_e(t)$ and $\pa(x)$. 
The only non-zero Poisson brackets are
\bes
[\pnu(s),\qmu(t)] &=& \dlt^\mu_\nu \dlt(s-t), \nl
{[} \pi_e(s), e(t)] &=& \dlt(s-t), 
\label{PB} \\
{[}\pa(x),\fb(y)] &=& -(-)^{\al\bt}[\fb(y), \pa(x)]
= \dlt^\al_\bt \dlt^N(x-y),
\eens
where $(-)^\al = 1$ ($(-)^\al = -1$) if $\fa$ is bosonic 
(fermionic) and $(-)^{\al\bt} = (-)^\al(-)^\bt$; the trajectory
and einbein are both bosonic.
Denote by $\CF(\QQ)$ and $\CF(\PP)$ the spaces of local functionals
over $\QQ$ and $\PP$; (anti)symmetri\-zation is automatically taken
into account by the bosonic (fermionic) character of the fields.
Then
\bes
\Lxi &=& \int dt\ \xmu(q(t))\pmu(t) -\nl
 &&- \int \dNx (\xmu(x)\dmu\fa(x) 
 + \dnu\xmu(x)\rep^\bt_\al(T^\nu_\mu)\fb(x)) \pa(x)
\label{class}\\
L_f &=& \int dt\ f(t)(-\dot\qmu(t)\pmu(t) +  e(t)\dot\pi_e(t)),
\eens
defines an embedding $DRO(N) \into \CF(\PP)$. 

Consequently, (\ref{class}) defines a $DRO(N)$ realization (by 
graded Poisson brackets) on $\CF(\QQ)$ and $\CF(\PP)$. Explicitly,
\bes
{[}\Lxi,\fa(x)] &=& -\xmu(x)\dmu\fa(x) 
 - \dnu\xmu(x)\rep^\bt_\al(T^\nu_\mu) \fb(x), \nl
{[}L_f,\fa(x)] &=& 0, \nl
{[}\Lxi,\qmu(t)] &=& \xmu(q(t)), \nl
{[}L_f,\qmu(t)] &=& -f(t)\dot\qmu(t),
\label{Lphi}\\
{[}\Lxi, e(t)] &=& 0, \nl
{[}L_f, e(t)] &=& -f(t)\dot e(t) - \dot f(t) e(t),
\nnl
{[}\Lxi,\pa(x)] &=&  -\xmu(x)\dmu \pa(x) 
 + \dnu\xmu(x)\pb(x) \rep^\al_\bt(T^\nu_\mu), \nl
{[}L_f,\pa(x)] &=& 0, \nl
{[}\Lxi,\pnu(t)] &=& -\dnu\xmu(q(t)) \pmu(t), \nl
{[}L_f,\pnu(t)] &=& -f(t)\dot\pnu(t) - \dot f(t) \pnu(t), 
\label{Lpi} \\
{[}\Lxi, \pi_e(t)] &=& 0, \nl
{[}L_f, \pi_e(t)] &=& -f(t)\dot\pi_e(t).
\eens

\subsection{ Euler-Lagrange constraint }
Let 
\be
S[\phi] = \int \dNx \sqg \LL(x;\phi)
\label{action}
\ee
be an action invariant 
under $diff(N)$, where the Lagrangian $\LL(x; \phi)$ is a $diff(N)$
scalar field of weight zero (not to be confused with the Lie derivative
$\Lxi$).
The notation emphasizes that the Lagrangian is a local functional of
$\phi$, i.e. a function of $\fa(x)$ and finitely many derivatives
at the spacetime point $x$. Moreover,
$|g(x)| = \eps^{\mu_1\mu_2..\mu_N}\eps^{\nu_1\nu_2..\nu_N}
g_{\mu_1\nu_1}(x)g_{\mu_2\nu_2}(x)\ldots g_{\mu_N\nu_N}(x)$
is the determinant of the metric, although the only important point 
is that $\sqg$ has weight one.

The solutions to the Euler-Lagrange (EL) equations,
\be
\Ea(x; \phi) \equiv [\pa(x), S] \equiv {\dlt S\/\dlt\fa(x)} = 0,
\label{EL}
\ee
define the {\em stationary surface} $\Sigma \subset \QQ$. 
The EL equations also generate the multiplicative ideal
$\NS = \{ f_\al\Ea(x; \phi) : f_\al\in\CF(\QQ) \} 
\subset \CF(\QQ)$. 
The factor space $\CF(\QQ)/\NS$ can be identified with the algebra 
of local functionals on the stationary surface, i.e. $\CF(\Sigma)$.
Conversely, $\Sigma$ 
can be recovered as the set of maximal ideals of $\CF(\Sigma)$, so
knowledge of this algebra is equivalent to solving the EL equations.
The problem is now to describe $\CF(\Sigma)= \CF(\QQ)/\NS$ in a 
simple manner. This space admits a resolution in terms of a certain
Koszul-Tate (KT) complex \cite{HT92};
the present exposition was mainly inspired by Stasheff \cite{Sta97}.
Recall that a complex $\Omega^\bullet$ is a collection of spaces 
$\Omega^g$ and maps $\dlt_g$,
\be
\ldots \map{\dlt_{-3}} \Omega^{-2}  
\map {\dlt_{-2}} \Omega^{-1} 
\map {\dlt_{-1}} \Omega^0 
\map {\dlt_0} \Omega^1 
\map {\dlt_1} \Omega^2 \ldots,
\label{complex}
\ee
such that $\dlt_g\dlt_{g-1} = 0$. The cohomology spaces are
$H^g(\dlt) = \ker \dlt_g/\im \dlt_{g-1}$. 
The complex (\ref{complex}) yields a one-sided resolution of a space 
$V$ if $H^\ell(\dlt) = V$, $\Omega^g = 0$ if $g>\ell$,
and $H^g(\dlt) = 0$ if $g<\ell$.

For each component of the EL equation $\Ea(x)$ (the functional 
dependence on $\phi$ is henceforth suppressed), 
introduce an {\em antifield} $\fsa(x)$ with
Grassmann parity opposite to $\fa(x)$ (and $\Ea(x)$).
Assign ghost numbers $\gh\,\fa(x) = 0$, $\gh\,\fsa(x) = -1$.
The term ``ghost number'' is perhaps somewhat misleading, since I never
introduce any ghosts, but the name is chosen in analogy with the 
Batalin-Vilkovisky terminology, see subsection \ref{ssec:BV} below.
The KT complex is the space $\Omb = \CF(\QQ) \otimes \CC[b]$, 
where the second factor consists of local polynomial functionals 
in the antifields; (anti-)symmetrization is automatically taken care of
by the (anti-)commuting nature of the antifields. 
This complex is naturally graded by ghost number,
and there is a nilpotent KT differential $\dlt$, defined by
\be
\dlt \fa(x) = 0, \qquad \dlt \fsa(x) = \Ea(x).
\ee
Since $\ker\dlt_0 = \CF(\QQ)$ and  $\im\dlt_{-1} = \NS$, 
$H^0(\dlt) = \CF(\QQ)/\NS = \CF(\Sigma)$ as desired.
Moreover, in the absense of Noether identities, $H^g(\dlt) = 0$, 
$g<0$, and $\Omega^g = 0$, $g>0$, so we
have obtained a resolution of $\CF(\Sigma)$.

For each antifield $\fsa(x)$, introduce an antifield momentum 
$\psa(x)$ satisfying Poisson brackets
\be
[\psa(x), \fsb(y)] = \dlt^\bt_\al \dlt^N(x-y).
\ee
Since the antifield has opposite Grassman parity compared to $\fa(x)$,
this bracket is symmetric if the original field is bosonic and vice versa.
The KT differential can now be expressed as 
\be
\dlt f = [\QKT, f], \qquad \forall f \in \CF(\QQ) \otimes \CC[b],
\label{dKT}
\ee
where the fermionic KT generator $\QKT$ is 
\be
\QKT = \int \dNx \Ea(x) \psa(x).
\label{QKT1}
\ee
$[\QKT,\QKT] = 0$ because $\Ea(x)$ commutes with 
the antifield momentum. 
Formula (\ref{dKT}) extends $\dlt$ to the phase space analogue of 
the KT complex, $\Omb = \CF(\PP) \otimes \CC[\phi^*, \pi^*]$ 
(polynomial functionals in antifields and antifield momenta): 
\be
\dlt \pa(x) = - \int \dNy (-)^\al {\dlt\Eb(y)\/\dlt\fa(x)} 
 \psb(y),
\qquad \dlt \psa(x) = 0,
\ee
where ${\dlt\Eb(y)/\dlt\fa(x)} = [\pa(x), \Eb(y)]$.

Before evaluating the cohomology, we note that $\Omb$ admits a double
grading. Assign ghost numbers by
\bes
&&\gh\,\fsa(x) = -1, \qquad
\gh\,\psa(x) = +1, 
\label{ghnum} \\
&&\gh\,\fa(x) = \gh\,\pa(x) = \gh\,\qmu(t) = \gh\,\pmu(t) =
\gh\,e(t) = \gh\,\pi_e(t) = 0.
\eens
We have $\gh\,[f,g] = \gh\,f + \gh\,g$ and $\gh\,Q = +1$. We may write
$\gh\,f = [N_\gh, f]$, where the ghost number operator 
$N_\gh\,= -\int \dNx \fsa(x)\psa(x)$.
Moreover, introduce the {\em momentum number} by
\bes
&&\mom\, \pa(x) = \mom\, \psa(x) = \mom\, \pmu(t) 
 = \mom\, \pi_e(t) = 1, 
\nllbnl{momnum}
&&\mom\, \fa(x) = \mom\, \fsa(x) = \mom\, \qmu(t) 
 = \mom\, e(t) = 0.
\eens
We have $\mom\, [f,g] = \mom\, f + \mom\, g - 1$ and $\mom\, Q = +1$, but
contrary to $\gh$, $\mom\,$ can not be expressed in bracket form.

$DRO(N)$ acts as follows on $\Omb$:
$\Ea(x)$ transforms as $\pa(x)$, the antifields 
are defined to transform in the same way, and thus the 
antifield momenta behave like $\fa(x)$.
\bes
{[}\Lxi, \Ea(x)] &=&  -\xmu(x)\dmu \Ea(x) 
 + \dnu\xmu(x)\Eb(x) \rep^\al_\bt(T^\nu_\mu), \nl
{[}\Lxi, \fsa(x)] &=&  -\xmu(x)\dmu \fsa(x) 
 + \dnu\xmu(x)\fsb(x) \rep^\al_\bt(T^\nu_\mu), 
\nllbnl{Lfp}
{[}\Lxi,\psa(x)] &=& -\xmu(x)\dmu \psa(x) 
 - \dnu\xmu(x)\rep^\bt_\al(T^\nu_\mu) \psb(x), \nl
{[}L_f, \Ea(x)] &=& [L_f, \fsa(x)] = [L_f, \psa(x)] = 0.
\eens
Hence the antifield contribution to the $DRO(N)$ generators is
\bes
\Lxi^{(1)} &=& \int \dNx ( -\xmu(x)\dmu \fsa(x) 
 + \dnu\xmu(x)\fsb(x) \rep^\al_\bt(T^\nu_\mu) ) \psa(x), \nl
L_f^{(1)} &=& 0.
\ees
$\gh\,\Lxi = \gh\,L_f = 0$ and $\mom\, \Lxi = \mom\, L_f = 1$,
which means that $DRO(N)$ commutes with the KT generator $\QKT$ and
the momentum number is preserved.

The KT complex has the double decomposition
\be
\Omb = \sum_{g=-\infty}^\infty \sum_{\ell=-\infty}^g \Omega^g_\ell,
\ee
where $\gh\,\Omega^g_\ell = g$, $\mom\, \Omega^g_\ell = \ell$.
Since the ghost and momentum numbers are preserved, each cohomology 
group $H^g_\ell(\QKT)$ is separately a $DRO(N)$ module. The
case $\ell=0$ was described above: $H^g_0(\QKT) = \dlt^g_0 \CF(\QQ)/\NS$.
Similarly, $H^g_\ell(\QKT) = 0$ if $g<0$, and
$H^\ell_\ell(\QKT)$ is the space of local functionals of $\phi(x)$ and 
$\psa(x)$ of the form
\be
f^{\al_1..\al_\ell}(\phi) \pi^*_{\al_1}\ldots \pi^*_{\al_\ell},
\label{lform}
\ee
modulo the ideal generated by relations
\be
\Ea(x) = 0 \qquad \hbox{and}\qquad
-\int \dNy (-)^\al {\dlt\Eb(y)\/\dlt\fa(x)}\psb(y) = 0.
\label{Epi}
\ee
The expression (\ref{lform}) is recognized as an $\ell$-form over $\QQ$. 
Since the antifield momenta commute with the fields and anticommute 
among themselves (for bosonic degrees of freedom), they can be thought
of as differentials; schematically, $\psa = d\fa$.
We have thus obtained a resolution of the space of $\ell$-forms on
the stationary surface $\Sigma$, and hence another description of 
$\Sigma$ itself. 

\subsection{ Auxiliary fields }
There is considerable freedom to describe the cohomology spaces in
non-minimal ways, by introducing auxiliary fields that are completely 
specified by their EL equations. 
An action the form 
$S^{(1)}[\phi,\psi]$, where $\psi_A(x) \equiv f_A(x;\phi)$,
gives rise to the same KT cohomology as
\be
S^{(2)}[\phi,\psi,\la] = S^{(1)}[\phi,\psi] 
 +\int \dNx  \la^A(x)(\psi_A(x) - f_A(x; \phi)),
\ee
where $\psi_A$ is treated as an independent field and
$\la^A$ is a Lagrangian multiplier field.
The EL equations for $\la^A$ and $\psi_A$,
\be
\psi_A(x) = f_A(x; \phi), \qquad
\la^A(x) = - {\dlt S^{(1)}\/\dlt \psi_A(x)},
\label{psila}
\ee
leave the same EL equations for $\fa$:
\be
{\dlt S^{(2)}\/\dlt \fa(x)} =
{\dlt S^{(1)}\/\dlt \fa(x)} + \int \dNy 
{\dlt S^{(1)}\/\dlt \psi_A(y)} {\dlt f_A(y)\/\dlt\fa(x)} = 0,
\ee
where (\ref{psila}) was used in the first step.
Therefore, the cohomologies defined by the KT generators
\bes
\QKT^{(1)} &=& \int \dNx ( {\dlt S^{(1)}\/\dlt\fa(x)} +
 \int \dNy {\dlt S^{(1)}\/\dlt \psi_A(y)} {\dlt f_A(y)\/\dlt\fa(x)}
 ) \psa(x), \nl
\QKT^{(2)} &=& \int \dNx \Big(
 ({\dlt S^{(1)}\/\dlt\fa(x)} 
 - \int \dNy \la^A(y) {\dlt f_A(y)\/\dlt\fa(x)} )\psa(x) +\\
 &&+ (\la^A(x)-{\dlt S^{(1)}\/\dlt\psi_A(x)}){\dlt\/\dlt\psi^{*A}(x)}
 + (\psi_A(x)-f_A(x)){\dlt\/\dlt\la^*_A(x)} \Big),
\eens
are identical. 
Here $\dlt/\dlt\psi^{*A}$ and $\dlt/\dlt\la^*_A$
are the antifield momenta corresponding to $\psi_A$ and $\la^A$,
respectively.

In the main cases of physical interest,
the KT generator can be made polynomial in all fields, provided that
sufficiently many auxiliary fields are included.
The following examples define some auxiliary fields that are needed 
below. Henceforth, they are tacitly assumed to be eliminated in 
cohomology by their EL equations.

1. The metric field $g_{\mu\nu}(x)$ has an inverse $g^{\mu\nu}(x)$,
which can be regarded as an auxiliary field obeying the equation
\be
g_{\mu\rho}(x)g^{\rho\nu}(x) = \dlt^\nu_\mu.
\ee

2. The weight one field $\upsilon(x) = \sqg$ used to densitize the 
Lagrangian can be eliminated by $\upsilon(x)^2 = |g(x)|$.

3. The Levi-Civit\`a connection is given by the usual formula
\be
\Gamma^\nu_{\si\tau}(x) = \half g^{\nu\rho}(x)(\dsi g_{\rho\tau}(x)
+ \dtau g_{\si\rho}(x) - \drho g_{\si\tau}(x)). 
\ee
It verifies
\bes
[\Lxi, \Gamma^\nu_{\si\tau}(x)] &=& 
 -\xmu(x)\dmu\Gamma^\nu_{\si\tau}(x) 
 + \drho\xnu(x)\Gamma^\rho_{\si\tau}(x) \nl
&& -\dsi\xmu(x)\Gamma^\nu_{\mu\tau}(x)
- \dtau\xmu(x)\Gamma^\nu_{\si\mu}(x) 
- \dsi\dtau\xnu(x), \nl
{[}L_f, \Gamma^\nu_{\si\tau}(x)] &=& 0.
\label{LC}
\ees
We can now define the covariant (w.r.t. diffeomorphisms) derivative
\be
\nabla_\mu = \dmu - \Gamma^\rho_{\nu\mu}(x)\rep(T^\nu_\rho).
\ee

4. The inverse of the einbein $e^{-1}(t)$, defined by 
$e^{-1}(t)e(t) = 1$.

5. The repara\-metrization connection $\Gamma(t) = -e^{-1}(t)\dot e(t)$,
transforming as
\bes
{[}L_f, \Gamma(t)] &=& -\dot f(t) \Gamma(t) 
 - f(t)\dot\Gamma(t) + \ddot f(t), 
\nllbnl{Virconn}
{[}\Lxi,\Gamma(t)] &=& 0.
\eens
Just as the Levi-Civit\`a connection can be used to define a 
derivative which is covariant w.r.t. $diff(N)$, $\Gamma(t)$ is needed
to define a derivative covariant w.r.t. repara\-metrizations $diff(1)$.

\subsection{ Geodesic constraint }
\label{ex:geo}
Just as the fields are restricted to Cauchy data by the EL equations, 
the observer's
trajectory can be eliminated by the geodesic equation, and the einbein
is an auxiliary field satisfying
\be
e(t) = \sqrt{g_{\mu\nu}(q(t)) \dot\qmu(t)\dot\qnu(t)}.
\ee
These equations can also be cast in EL form. 
Add to (\ref{action}) the term
\be
S^{(q)}[q,e,g] = -\half \int dt\ e(t) +
 e^{-1}(t) g_{\mu\nu}(q(t)) \dot\qmu(t)\dot\qnu(t),
\label{Sq}
\ee
so the total action is $S[\phi,q,e] = S[\phi] + S^{(q)}[q,e,g]$.
Note that $S^{(q)}$ depends on the metric, which is included in the
set of fields.
Define
\bes
\GG_\mu(t) &\equiv& [\pmu(t), S] \nl
&=& e^{-1}(t) g_{\mu\nu}(q(t)) ( \ddot\qnu(t) + \Gamma(t) \dot\qnu(t) 
+ \Gamma^\nu_{\si\tau}(q(t))\dot\qsi(t)\dot\qtau(t) ), \nl
\OO(t) &\equiv& [\pi_e(t), S] 
\label{geo} \\
&=&  \half( e^{-2}(t) g_{\mu\nu}(q(t)) \dot\qmu(t)\dot\qnu(t) - 1),
\eens
where $\Gamma(t)$ is the repara\-metrization connection (\ref{Virconn}).
These operators transform homogeneously:
\bes
{[}\Lxi, \GG_\nu(t)] &=& -\dnu\xmu(q(t))\GG_\mu(t), \\
{[}L_f, \GG_\nu(t)] &=& 
 -f(t)\dot\GG_\nu(t) - \dot f(t)\GG_\nu(t),
\nnl
{[}\Lxi, \OO(t)] &=& 0, \\
{[}L_f,  \OO(t)] &=& -f(t)\dot\OO(t).
\eens
We now introduce the trajectory antifield $q^*_\mu(t)$, with momentum
$p^{*\mu}(t)$, and the einbein antifield $e^*(t)$, with momentum
$\pi^*_e(t)$. Since $\GG_\nu(t)$ and $\OO(t)$ are bosonic, these
antifields are fermionic and obey the non-zero anticommutation relations
\be
[p^{*\mu}(s), q^*_\nu(t)] = \dlt^\mu_\nu \dlt(s-t), \qquad
[\pi^*_e(s), e^*(t)] = \dlt(s-t).
\ee
Ghost and momentum numbers are given by by
\bes
\gh\,q^*_\nu(t) = \gh\,e^*(t) = -1,
&\quad&
\gh\,p^{*\mu}(t) = \gh\,\pi^*_e(t) = +1, \\
\mom\,q^*_\nu(t) = \mom\,e^*(t) = 0,
&\quad&
\mom\,p^{*\mu}(t) = \mom\,\pi^*_e(t) = 1,
\eens
By adding the term 
\be
\QKT^{(\GG)} = \int dt\ \GG_\mu(t) p^{*\mu}(t) + \OO(t)\pi^*_e(t)
\label{QGG}
\ee
to the KT differential, the constraints 
$\GG_\nu(t) \approx \OO(t) \approx 0$
are implemented in cohomology.

\subsection{ Noether identities }
The previous discussion ignored the possibility of relations between
the EL equations. This is certainly incorrect; at the very least,
the DRO algebra imply certain conditions. In general we assume that
there are Noether identities of the form
\be
r^a(x) \equiv \int \dNy r^a_\al(x,y)\Ea(y)
= \int \dNy (-)^\al \Ea(y)r^a_\al(x,y) \equiv 0.
\label{Noether}
\ee
For simplicity, let all Noether identities be independent;
the addition of non-trivial relations between them is straightforward
but leads to unnecessary complications. For each Noether identity,
introduce a Noether (or second-order) antifield $\bb^a(x)$ 
with momentum $\cc_a(x)$. 
We only deal with bosonic Noether identities, and require their
antifields to be bosonic as well. The non-zero Poisson bracket is
\be
[\cc_a(x), \bb^b(y)] = \dlt^b_a \dlt^N(x-y).
\label{bc}
\ee
A new (fermionic) term has to be added to the KT generator (\ref{QKT1})
\be
\QKT^{(2)} = \iint \dNx \dNy (-)^\al r^a_\al(x,y) \fsa(y) \cc_a(x).
\label{Q2}
\ee
The modified KT differential acts as
\bes
{[}\QKT, \fa(x)] &=& 0, \nl
{[}\QKT, \fsa(x)] &=& \Ea(x), 
\label{dKT2} \\
{[}\QKT, \bb^a(x)] &=& \int \dNy (-)^\al r^a_\al(x,y)\fsa(y),
\nnl
{[}\QKT, \pa(x)] &=& - \int \dNy 
 (-)^\al {\dlt \Eb(y)\/\dlt\fa(x)} \psb(y) -\nl
&&-\int \dNy\dNz 
(-)^{\al+\bt}{\dlt r^a_\bt(y,z)\/\dlt\fa(x)} \fsb(z) \cc_a(y), \nl
{[}\QKT, \psa(x)] &=& \int \dNy r^a_\al(y,x) \cc_a(y), \\
{[}\QKT, \cc_a(x)] &=& 0,
\eens
It follows from (\ref{Noether}) that the KT generator is still nilpotent:
\be
{[}\QKT, \QKT] = 2\iint \dNx \dNy 
 (-)^\al \Ea(x) r^a_\al(x,y)\cc_a(y)
\equiv 0.
\ee
The addition of Noether antifields is necessary because we want the
KT complex to yield a resolution, i.e. $H^g_\ell(\QKT) = 0$ if 
$g\neq\ell$. In the presence of Noether identities,
\be
\dlt \int \dNy (-)^\al r^a_\al(x,y)\fsa(y) 
= \int \dNy r^a_\al(x,y) \Ea(y) \equiv 0,
\ee
so $\ker \dlt_{-1} \neq 0$, but this expression is exact by
(\ref{dKT2}), so $H^{-1}(\QKT)$ still vanishes.

\subsection{ Gauge symmetries }
As is well known, Noether identities are connected to gauge symmetries.
{F}rom (\ref{EL}) and (\ref{Noether}) immediately follows that
\be
\J_X = \iint \dNx \dNy X_a(x) r^a_\al(x,y) \pa(y) 
\label{JX}
\ee
satisfies $[\J_X, S] = 0$. The set of such operators generate a
Lie algebra, which is easily seen as follows. If $[\J_X, S] = 
[\J_Y, S] = 0$, $[[\J_X, \J_Y], S] = 0$ by the Jacobi identities.
If some Noether identity were fermionic, (\ref{JX}) would
define a super-Lie algebra, but this possibility is not considered here.
Note that we use the same notation as for the proper gauge algebra
$map(N,\oj)$, but the present exposition is more general; in particular,
it includes $DRO(N)$. This overloading should not cause confusion.

Assume that the Noether algebra can be written in localized form as
\be
[\J^a(x), \J^b(y)] = \int \dNz f^{ab}{}_c(x,y;z) \J^c(z),
\ee
where 
\be
\J^a(x) = \int \dNy r^a_\al(x,y) \pa(y)
\ee
and the structure constants $f^{ab}{}_c(x,y;z)$ depends on
(finite derivatives of) $\dlt(x-z)$ and $\dlt(y-z)$ only. This is an
assumption about locality which is always valid. Then the following
identity holds
\bes
&&\int \dNz r^a_\al(x,z) {\dlt r^b_\bt(y,w)\/\fa(z)}
- r^b_\al(y,z) {\dlt r^a_\bt(x,w)\/\fa(z)} \nl
&&= \int \dNz f^{ab}{}_c(x,y;z)r^c_\bt(z,w).
\ees

The action of $\J_X$ on the antifields is fixed by 
demanding that $[\J_X,\QKT] = 0$. We find
\bes
[\J_X, \fa(x)] &=& \int \dNy X_a(y) r^a_\al(y,x), \nl
{[}\J_X, \fsa(x)] &=& -\iint \dNy\dNz (-)^{\al\bt+\bt} 
 X_a(y) \fsb(z){\dlt r^a_\bt(y,z)\/\dlt\fa(x)}, \\
{[}\J_X, \bb^a(x)] &=& - \iint \dNy\dNz
 f^{ab}{}_c(x,y;z) X_b(y) \bb^c(z),
\nnl
{[}\J_X, \pa(x)] &=& 
-\iint \dNy\dNz  X_a(y) {\dlt r^a_\bt(y,z)\/\dlt\fa(x)}\pb(z), \nl
{[}\J_X, \psa(x)] &=&  \iint \dNy\dNz (-)^{\al\bt+\al} 
X_a(y) {\dlt r^a_\al(y,x)\/\dlt\fb(z)} \psb(z), \\
{[}\J_X, \cc_a(x)] &=& \iint \dNy\dNz
 f^{cb}{}_a(z,y;x) X_b(y) \cc_c(z).
\eens
The total generators are thus 
$\J_X^\tot = \J_X + \J_X^{(1)} + \J_X^{(2)}$, where
\bes
\J_X^{(1)} &=& -\iiint \dNx\dNy\dNz (-)^{\al\bt+\bt} 
 X_a(y) \fsb(z) {\dlt r^a_\bt(y,z)\/\dlt\fa(x)} \psa(x), \nl
\J_X^{(2)} &=& - \iiint \dNx \dNy\dNz
 f^{ab}{}_c(x,y;z) X_b(y) \bb^c(z) \cc_a(x).
\ees

The Noether identity (\ref{Noether}) can be rewritten as
\be
\int \dNx [\J_X, \fa(x)] \Ea(x) \equiv 0.
\label{JXNoet}
\ee
Hence not only do Noether identities imply local symmetries, but the
converse is also true. Note that the bosonic character of the Noether
identities is manifest here.
In particular, diffeomorphism symmetry implies
\be
\int \dNx [\Lxi, \fa(x)] \Ea(x) 
+ \int dt\ [\Lxi, \qmu(t)] \GG_\mu(t) = 0
\label{LxiNoet}
\ee
($diff(N)$ acts trivially on the einbein),
and repara\-metrization symmetry gives
\be
\int dt\ [L_f, \qmu(t)] \GG_\mu(t) +
\int dt\ [L_f, e(t)] \OO(t) \equiv 0
\label{LfNoet}
\ee
($diff(1)$ acts trivially on the fields).
The corresponding additions to the KT generator are
\bes
\QKT^{(\diff)} &=& \int \dNx \Big(
 \int \dNy (-)^\al[\LL_\mu(x), \fa(y)] \fsa(y) +\nl
 &&+ \int dt\ [\LL_\mu(x), \qnu(t)] q^*_\nu(t) \Big)\cc^\mu(x), \nl
\QKT^{(\rrep)} &=&  
\iint dsdt\ ( [L(s), \qmu(t)] q^*_\mu(t) + [L(s), e(t)] e^*(t)) \cc(s), 
\label{QNoet} \\
\QKT^{(\gauge)} &=& \iint \dNx\dNy
 (-)^\al [\J^a(x), \fa(y)] \fsa(y)\cc_a(x),
\eens
where the localized generators were defined in (\ref{local}), 
the Noether antifields are $\bb_\mu(x)$, $\bb(t)$, $\bb^a(x)$,
and their momenta are $\cc^\mu(x)$, $\cc(t)$ and $\cc_a(t)$, 
respectively.

The gauge algebra needs only be satisfied up to a KT exact term.
For every choice of fermionic operator $K_X$, the modified generators
$\J'_X = \J_X + [\QKT, K_X]$ satisfy the same algebra in cohomology
as does the original $\J_X$, although the brackets on $\Omb$ acquires
a correction:
\be
[\J'_X, \J'_Y] = \J'_{[X,Y]} + [\QKT, [\J_X, K_Y]
 - [\J_Y, K_X] - K_{[X,Y]}].
\ee
However, this freedom will not be exploited further.
     
\subsection{ Continuity equation }
\label{sec:conti}
It often happens that the fields can be split into two disjoint sets,
$\fa = (\varphi_i, \psi_A)$, such that the action takes the form
$S = S_1[\varphi] + S_2[\varphi,\psi]$. Typically, $\varphi_i$ is a
metric or gauge field, and $\psi_A$ denote matter fields.
Moreover, we demand that
the Noether symmetries commute with each piece separately, i.e.
$[\J_X, S_1] = [\J_X, S_2] = 0$.
Then there are two independent identities
\bes
&& \int \dNy r^a_i(x,y) {\dlt S_1\/\dlt\varphi_i(y)} \equiv 0, 
\label{NI1} \\
&& \int \dNy r^a_i(x,y) {\dlt S_2\/\dlt\varphi_i(y)} 
+ r^a_A(x,y) {\dlt S_2\/\dlt\psi_A(y)} \equiv 0.
\label{NI2}
\ees
However, these are not separately proportional to the EL equations
\be
{\dlt S_1\/\dlt\varphi_i(x)} + {\dlt S_2\/\dlt\varphi_i(x)} = 0, \qquad
{\dlt S_2\/\dlt\psi_A(x)} = 0,
\ee
so only the sum of (\ref{NI1}) and (\ref{NI2}) imposes restrictions
on the EL equations, provided, of course, that $S_2$ depends non-trivially 
on $\varphi_i$. Combining (\ref{NI1}) and the EL equations we find
\be
\int \dNy r^a_i(x,y) {\dlt S_2\/\dlt\varphi_i(y)} = 0
\ee
on the stationary surface. This is the continuity equation. However, it
is not an identity that holds off the stationary surface, and therefore
it needs not be eliminated in cohomology.

\subsection{ Examples }
\subsubsection{ Maxwell-Dirac }
The fields $\fa$ consist of  the bosonic gauge potential $A_\mu$ 
and two independent fermionic Dirac spinors $\psi$ and $\bar\psi$.
The spacetime points $x$ and spinor indices are suppressed in this
example, and fermionic brackets are explicitly indicated by 
$\{\cdot,\cdot\}$. 
For brevity, we assume the metric to be flat Minkowski, 
and freely use this metric to raise and lower indices, and hence
$diff(N)$ is broken down to the Poincar\'e algebra. Let $\gm_\mu$
denote gamma matrices, $\{\gm_\mu,\gm_\nu\} = 2g_{\mu\nu}$.
Henceforth, we focus on the $map(N,u(1))$ Noether symmetry.

The action reads
\be
S = -{1\/4} \int F_{\mu\nu}F^{\mu\nu} 
 + \int \bar\psi(\gm^\mu(\dmu + iA_\mu) - m) \psi,
\ee
where the field strength $F_{\mu\nu} = \dmu A_\nu - \dnu A_\mu$.
According to our prescription, we introduce antifields and momenta
as follows.
\[
\begin{array}{lllll}
\hbox{Fields} &  \fa & A_\mu & \psi & \bar\psi\sim \psi^\dagger\gm^0 \\
\hbox{Momenta} & \pa & E^\mu & \pi\sim\psi^\dagger 
 & \bar\pi\sim\bar\psi^\dagger \\
\hbox{Antifields} & \fsa & A^{*\mu} & \psi^*\sim\psi^\dagger
 & \psi^*\sim\bar\psi^\dagger \\
\hbox{Antifield momenta} & \psa & E^*_\mu & \pi^*\sim\psi 
&\bar\pi^*\sim\bar\psi, \\
\end{array}
\]
where $\sim$ indicates the transformation properties under rotations.
The non-zero Poisson brackets are
\bes
[E^\mu, A_\nu] = \dlt^\mu_\nu, 
&\qquad&
\{\pi, \psi\} = \{\bar\pi, \bar\psi\} = 1, \nle
\{E^*_\mu, A^{*\nu}\} = \dlt^\nu_\mu,
&\qquad&
[\pi^*, \psi^*] = [\bar\pi^*, \bar\psi^*] = 1.
\eens
The EL equations are
\bes
{\dlt S\/\dlt A_\mu} &\equiv& [E^\mu,S] 
= \dnu F^{\nu\mu} - j^\mu, \nl
j^\mu &\equiv& \bar\psi\gm^\mu\psi, \nle
{\dlt S\/\dlt\psi} &\equiv& [\pi,S]
= (i\dmu + A_\mu) \bar\psi\gm^\mu + m\bar\psi, \nl
{\dlt S\/\dlt\bar\psi} &=& [\bar\pi,S]
= \gm^\mu(i\dmu - A_\mu)\psi - m\psi.
\eens
The first part of the KT generator is
\bes
\QKT^{(1)} &\equiv& \int \Ea\psa 
 =\int (\dnu F^{\nu\mu} - j^\mu)E^*_\mu +\\
&& + ((i\dmu + A_\mu) \bar\psi\gm^\mu + m\bar\psi)\pi^*
 - \bar\pi^*(\gm^\mu(i\dmu - A_\mu)\psi - m\psi).
\eens
The Noether identity reads
\bes
&& -\dmu{\dlt S\/\dlt A_\mu} + i\bar\psi{\dlt S\/\dlt\bar\psi}
+ i{\dlt S\/\dlt\psi}\psi 
= -\dmu(\dnu F^{\nu\mu} - j^\mu) +
\label{u1N} \\
\bl+i\bar\psi(\gm^\mu(i\dmu - A_\mu)\psi - m\psi)
+i((i\dmu + A_\mu) \bar\psi\gm^\mu + m\bar\psi)\psi
\equiv 0.
\eens
The corresponding gauge symmetry is $map(N, u(1))$,
which acts as follows on the fields
\be
[\J_X, A_\mu] = \dmu X, \qquad
[\J_X, \psi] = -iX\psi, \qquad
[\J_X, \bar\psi] = iX\bar\psi.
\ee
To eliminate this symmetry, we must introduce the Noether antifield
$\bb$, with momentum $\cc$: $[\cc,\bb] = 1$. They transform in the
adjoint representation of the gauge algebra, which in this case is
trivial since $u(1)$ is abelian:
$[\J_X, \bb] = [\J_X, \cc] = 0$.
The total gauge generator is
\be
\J_X = \int \dmu X E^\mu + iX (\pi\psi + \bar\psi\pi 
+\psi^*\pi^* + \bar\pi^*\bar\psi^*),
\ee
and the Noether contribution to the KT generator is
\be
\QKT^{(2)} = -\int (\dmu A^{*\mu} 
 + i\psi^*\psi + i\bar\psi\bar\psi^*)\cc.
\ee
In fact, (\ref{u1N}) is of the form discussed in subsection
(\ref{sec:conti}). The first Noether identity (\ref{NI1}) reads 
$\dmu\dnu F^{\mu\nu} \equiv 0$, leading to the continuity equation
$\dmu j^\mu = 0$.

\subsubsection{ Yang-Mills and spinors }
The example in the previous subsection can be extended to the 
Yang-Mills case, by replacing the gauge group $u(1)$ by an arbitrary
semi-simple Lie algebra $\oj$. The modifications are straightforward
and are left to the reader.

To describe spinors in a $diff(N)$ invariant manner requires a vielbein
formalism. This reduces
to the Yang-Mills case with gauge group $so(N)_{spin}$, except that 
we can define a vielbein $e^I_\mu(x)$ with inverse $e^{I\mu}(x)$
($I,J,\ldots$ denote $so(N)$ vector indices), such that the spin 
connection and the metric are auxiliary fields, satisfying
\be
\omega^{IJ}_\mu(x) =  e^{[I}_\nu(x)\dmu e^{J]\nu}(x), \qquad
g_{\mu\nu}(x) = e^I_\mu(x) e^I_\nu(x).
\ee

\subsubsection{ Einstein }
The action reads
\be
S = S^{(E)}[g] + S'[g,\phi] + S^{(q)}[q,e,g],
\ee
where the Einstein action is
\be
S^{(E)}[g] = {1\/16\pi} \int \dNx \sqg R(x),
\ee
and $R(x)$ is the scalar curvature.
Further, $S'[g,\phi]$ is the part of the action depending on other
fields and $S^{(q)}[q,e,g]$ was defined in (\ref{Sq}).
The EL equation reads
\bes
{\dlt S\/\dlt g_{\mu\nu}(x)} &=& -{1\/16\pi}\sqg \Big(
G^{\mu\nu}(x) - 8\pi T^{\mu\nu}(x) - 
\label{EEL}\\
&&- 8\pi \int dt\ e^{-1}(t) \dot\qmu(t)\dot\qnu(t) \dlt^N(x - q(t)) 
\Big) = 0,
\eens
where $G^{\mu\nu} = R^{\mu\nu} - (1/2)g^{\mu\nu}R$ 
is the Einstein tensor and 
$T^{\mu\nu} = (2/\sqrt{|g|})\ab\dlt S'/\dlt g_{\mu\nu}$
is the energy-momentum tensor. The last, non-standard, term in
(\ref{EEL}) describes how the massive observer curves spacetime around 
herself.

Let $\pi^{\mu\nu}(x)$ denote the momentum conjugate to $g_{\mu\nu}(x)$,
and let $g^{*\mu\nu}(x)$ and $\pi^*_{\mu\nu}(x)$ be the fermionic 
antifield and its momentum. The contribution to the KT generator is
\bes
\QKT^{(E)} &=& -{1\/16\pi} \int \dNx 
 \sqg (G^{\mu\nu}(x) - 8\pi T^{\mu\nu}(x))\pi^*_{\mu\nu}(x) +\nl
&&+ \half \int dt\ \sqrt{|g(q(t))|}\, 
 e^{-1}(t) \dot\qmu(t)\dot\qnu(t)\pi^*_{\mu\nu}(q(t)) .
\ees
The Noether symmetry $diff(N)$ is of the form discussed in
subsection (\ref{sec:conti}). The full identity depends on all fields,
but there is also the identity $\dnu G^{\mu\nu}(x) \equiv 0$, 
which leads to the continuity equation $\dnu T^{\mu\nu}(x) = 0$.

\subsubsection{Geodesic constraint}
Continues subsection (\ref{ex:geo}).
The $diff(1)$ identity (\ref{LfNoet}) becomes
\be
\dot\qmu(t)\GG_\mu(t) - e(t)\dot\OO(t) \equiv 0,
\ee
and the reparametrization contribution to the KT generator becomes
\bes
\QKT^{(\rrep)} &=& \int dt\ 
 (\dot\qmu(t)q^*_\mu(t) - e(t)\dot e^*(t))\cc(t).
\ees

\subsection{ Comparison with the Batalin-Vilkovisky formalism }
\label{ssec:BV}
Since the formulation of classical physics that has been developped
in the previous subsections is new, it makes sense to compare it with
other approaches. The closest resemblance is with the
antifield formalism of Batalin-Vilkovisky (BV), particularly
in the cohomological formulation of \cite{HT92, Sta97}.
Similarly to these authors, I impose the EL equation in the cohomology 
generated by the KT differential. 
However, there are three major differences. 

1. In the BV approach one considers a BRST complex rather than the
KT complex, i.e. Noether symmetries are eliminated by the introduction
of ghosts. This could be done in the present formalism as well.
For each Noether identity (\ref{Noether}), introduce a fermionic ghost 
$C_a(x)$ and a ghost momentum (or antighost) $B^a(x)$.
The BRST generator
\bes
Q_{BRST} &=& \int \dNx C_a(x) \J^a(x) +\nle
&&+ \half\iiint \dNx\dNy\dNz f^{ab}{}_c(x,y;z) 
  C_a(x) C_b(y) B^c(z) 
\eens
is nilpotent, and its cohomology identifies points on the same 
gauge orbits. The total generator $Q_{TOT} = \QKT + Q_{BRST}$ is also
nilpotent, and its cohomology consists of gauge-equivalence classes of
differential forms on the stationary surface. 

However, the BRST generator will not appear in this work. 
Classically, this
is a matter of taste; it is equivalent to view a space as a
$\oj$ module or to consider its equivalence classes under the $\oj$
action. However, quantization will in general introduce abelian 
extensions (``anomalies''), which ruin the nilpotency of the BRST 
generator. Therefore, we only consider the KT generator, which is not 
affected by anomalies.

2. Not only do I use fields and antifields, but 
also field and antifield momenta. There is thus 
already a graded Poisson structure, in terms of which an antibracket 
(a non-zero fermionic bracket between the fields and antifields) can be
defined. For any $f, g \in \CF(\QQ)\otimes\CC[\phi^*,\bb,C]$, set 
\bes
(f,g) &=& \int \dNx \Big( -(-)^\al ([f, \psa(x)][\pa(x), g]
+ [f, \pa(x)][\psa(x), g] )  +\nl
&&+ [f, B^a(x)][\cc_a(x), g]
+ [f, \cc_a(x)][B^a(x), g] \Big) \\
&=& -(-)^{(f+1)(g+1)}(g,f).
\eens
In particular,
\be
(\fa(x), \fsb(y)) = \dlt^\bt_\al \dlt^N(x-y), \qquad
(C_a(x), \bb^b(y)) = \dlt^b_a \dlt^N(x-y).
\ee
The KT differential on $\CF(\QQ)\otimes\CC[\phi^*,\bb,C]$ 
is now reproduced by  $\dlt f = (f,S_\tot)$,
where 
\be
S_\tot = S + \int \dNx\dNy \fsa(y) r^a_\al(x,y)C_a(x) 
\ee
is the total action. Nilpotency leads to the classical
master equation $(S_\tot,S_\tot)=0$. 
However, this definition of $\dlt$ can not
be extended to all of $\CF(\PP)\otimes\CC[\phi^*,\pi^*,\bb,\cc,C,B]$,
because $(\pa(x), S_\tot) = (\psa(x), S_\tot)= (B^a(x), S_\tot) = 0$. 
Hence in the BV formalism, the KT complex only gives a resolution of 
the space of functions on the solution surface, whereas the 
$\ell$-form spaces $H^\ell_\ell(\QKT)$ can only be resolved using
the more general expression (\ref{dKT}).

3. Momenta and velocitites are treated independently. Usually, they
are identified by the equation
\be
\pa(x) = {\d(\sqrt{|g|(x)} \LL(x;\phi))\/\d\d_0\phi(x)}.
\label{momdef}
\ee
This equation can be thought of as an extra constraint, from which
either $\pa(x)$ or $\d_0\fa(x)$ can be eliminated. However, this
additional condition gives rise to three significant problems:
First, it is not generally covariant, so there is little
hope to represent $DRO(N)$ on the factor space.
Second, it is a second class constraint, which can not be separated into
a first class constraint and a gauge fixation in a natural way.
Hence cohomological methods fail.
Third, it can not be formulated in jet space, since $\pa(x)$ can not
be expanded in a Taylor series around $q(t)$.
In view of these difficulties, velocities and momenta are kept as
independent objects. The price for this seems modest: the cohomology
groups contain the $\ell$-form spaces for non-zero $\ell$.

\subsection{Quantization}
Having formulated classical physics as the cohomology of the KT 
complex, we could now try to quantize it by reinterpreting the
Poisson brackets (\ref{PB}) as commutators. The strategy is thus
first to quantize and then recover the dynamics in cohomology.
However, this leads to three major difficulties.

1. The geodesic equation contains the metric at the observer's present
position, $g_{\mu\nu}(q(t))$. It is not clear what to do with this
object if both $g_{\mu\nu}(x)$ and $\qmu(t)$ are turned into operators.

2. There is no invariant time choice. Of course, we could make a Fourier
transformation w.r.t. $x^0$, and define the Fock vacuum to be annihilated
by negative energy modes, but such a decomposition is not invariant.
Therefore, it is not clear that the Fock space carries a $diff(N)$
representation.

3. Normal ordering of the generators (\ref{class}) is ill defined. More
precisely, central extensions proportional to the number of 
$x^0$-independent functions arise, but this number is infinite except
in one dimension.

These difficulties disappear if we expand the fields in a Taylor series
around the observer's present position.

\section{ Jet space quantization }

\subsection{ Jet space trajectories }
Let $\mm = (m_0, \ab m_1, \ab .., \ab m_{N-1})$, all $m_\mu\geq0$, be a 
multi-index of order
$|\mm| = \sum_{\mu=0}^{N-1} m_\mu$, let $\mmu$ be a unit vector in the 
$\mu$:th direction, and let $0$ be the empty multi-index of order zero.
Expand $\fa(x)$ in a power series around $\qmu(t)$.
\be
\fa(x) = \sum_{|\mm|\geq0} {1\/\mm!} \phi_{\al,\mm}(t)(x-q(t))^\mm,
\label{gen}
\ee
where $\mm! = m_0! m_1! .. m_{N-1}!$ and
\be
(x-q(t))^\mm = (x^0-q^0(t))^{m_0} (x^1-q^1(t))^{m_1} ..
 (x^{N-1}-q^{N-1}(t))^{m_{N-1}}.
\ee
Since the DGRO algebra acts on $\CF(\QQ)$, it also acts
on the infinite jet space $J^\infty\QQ$, with basis $(\fm\al(t),
\qmu(t), e(t))$, $t\in S^1$. 
The transformation law is described in (\ref{Lfn}) below. 
$DGRO(N,\oj)$ also acts on the $p$-jet spaces $J^p\QQ$,
$p$ finite, obtained by truncating to $|\mm|\leq p$. 
The realization on $J^p\QQ$ is non-linear in the 
trajectory $\qmu(t)$, so it must be interpreted as a linear 
representation on $\CF(J^p\QQ)$, or more restrictively as a 
representation on $J^p\QQ \otimes_{[q(t)]} Obs(N)$, where the observer 
algebra $Obs(N)$ was defined in section \ref{sec:DRO} and $\qmu(t)$
is identified in both factors.

The corresponding phase space $J^p\PP$ is obtained by adjoining 
to $J^p\QQ$ dual coordinates $(\pim\al(t), \pmu(t), \pi_e(t))$. 
The only non-zero brackets are
\bes
[\pmu(s), q^\nu(t)] &=& \dlt^\nu_\mu \dlt(s-t), \nl
{[}\pi_e(s), e(t)] &=& \dlt(s-t),
\label{Poisson} \\
{[}\pim\al(s), \phi_{\bt,\nn}(t)] &\equiv& 
-(-)^{\al\bt}[\phi_{\bt,\nn}(t), \pim\al(s)]
 = \dlt^\al_\bt \dlt^\mm_\nn \dlt(s-t).
\eens
Observe that the $\pim\al(t)$ are not the Taylor 
coefficients of $\pa(x)$, because the latter can not be expanded
in a power series in $(x-q(t))$. The reason is that $\dlt^N(x-y)$
can not be written as a double power series. 
The delta function does have the expansion
\be
\dlt^N(x-y) = \sum_{\mm\in\ZZ} (x-q(t))^\mm (y-q(t))^{-\mm-\one},
\ee
where $\one = (1, 1, \ldots, 1)$, but to use this expression in
(\ref{PB}), $\fa(x)$ and $\pa(x)$ must be expanded in 
formal Laurent (rather than power) series. Since such an assumption
leads to the type of infinities that we want to avoid, we simply
postulate no relation between $\pa(x)$ and $\pim\al(t)$.
It is now clear why (\ref{momdef}) has no jet space analogue:
$\pim\al(t)$ has an upper multi-index whereas any function of
$\fm\al(t)$ can only have multi-indices downstairs.

Define $T^\mm_\nn(\xi), J^\mm_\nn(X) \in \UU(\gloj)$ 
(universal envelopping algebra) by
\bes
T^\mm_\nn(\xi) &=& 
 {\nn\choose\mm} \d_{\nn-\mm+\nnu}\xmu T^\nu_\mu \nl
 &+& {\nn\choose\mm-\mmu} (1-\dlt^{\mm-\mmu}_\nn)
 \d_{\nn-\mm+\mmu}\xmu I, \\
J^\mm_\nn(X) &=&
 {\nn\choose\mm} \d_{\nn-\mm} X_a J^a.
\eens
where ${\nn\choose\mm} = \nn!/\mm!(\nn-\mm)!$
and $I$ is the unit element in $\UU(\gloj)$.
These objects satisfy
\bes
T^\mm_{\nn+\nnu}(\xi) &=& \dnu\xmu\dlt^\mm_{\nn+\mmu}I
 + T^\mm_\nn(\dnu\xi) + T^{\mm-\nnu}_\nn(\xi), \nl
T^\mm_0(\xi) &=& \dlt^\mm_0 \dnu\xmu T^\nu_\mu, \nl
\dnu T^\mm_\nn(\xi) &=& T^\mm_\nn(\dnu\xi),
\label{Tmn} \\
T^\mm_\nn([\xi,\eta]) &=& 
\xmu T^\mm_\nn(\dmu\eta) - \ynu T^\mm_\nn(\dnu\xi) \nl
&& + \summrn T^\rr_\nn(\xi) T^\mm_\rr(\eta) 
 - T^\rr_\nn(\eta) T^\mm_\rr(\xi),
\nnl
J^\mm_{\nn+\mmu}(X) &=& J^\mm_\nn(\dmu X) + J^{\mm-\mmu}_\nn(X), \nl
J^\mm_0(X) &=& \dlt^\mm_0 X_a J^a, \nl 
\dmu J^\mm_\nn(X) &=& J^\mm_\nn(\dmu X), 
\label{Jmn} \\
J^\mm_\nn([X,Y]) &=&  \summrn 
 J^\rr_\nn(X)J^\mm_\rr(Y) - J^\rr_\nn(Y)J^\mm_\rr(X), \nl
J^\mm_\nn(\xmu\dmu X) &=& \xmu J^\mm_\nn(\dmu X)
 + \summrn T^\rr_\nn(\xi)J^\mm_\rr(X) - J^\rr_\nn(X)T^\mm_\rr(\xi).
\eens
In particular, $T^\mm_\nn(\xi) = J^\mm_\nn(X) = 0$ if $|\mm|>|\nn|$.
Alternatively,  (\ref{Tmn}--\ref{Jmn}) could be taken as a recursive
definition of $T^\mm_\nn(\xi)$ and $J^\mm_\nn(X)$.
Every $\gloj$ representation $\rep$ on $V$ clearly gives a 
representation of these operators.

We can now write down the $DGRO(N,\oj)$ action on $J^p\PP$.
\bes
{[}\Lxi,\fn\al(t)] &=& 
 -\summ{|\nn|} \rep^\bt_\al(T^\mm_\nn(\xi(q(t)))) \fm\bt(t), \nl
{[}\J_X, \fn\al(t)] &=& 
 - \summ{|\nn|} \rep^\bt_\al(J^\mm_\nn(X(q(t)))) \fm\bt(t),
\label{Lfn} \\
{[}L_f,\fn\al(t)] &=& -f(t)\dot\fn\al(t) - \la\dot f(t)\fn\al(t), 
\nnl
{[}\Lxi,\qmu(t)] &=& \xmu(q(t)), \nl
{[}\J_X, \qmu(t)] &=& 0, \\ 
{[}L_f,\qmu(t)] &=& -f(t)\dot\qmu(t),
\nnl
{[}\Lxi, e(t)] &=& {[}\J_X, e(t)] = 0, \\
{[}L_f, e(t)] &=& -f(t)\dot e(t) - \dot f(t) e(t),
\nnl
{[}\Lxi,\pim\al(t)] &=&  
 \summn p \pin\bt(t) \rep^\al_\bt(T^\mm_\nn(\xi(q(t)))), \nl
{[}\J_X, \pim\al(t)] &=&  
 \summn p \pin\bt(t) \rep^\al_\bt(J^\mm_\nn(X(q(t)))), \\
{[}L_f,\pim\al(t)] &=& -f(t)\dot\pim\al(t) -(1-\la)\dot f(t) \pim\al(t), 
\nnl
{[}\Lxi,\pnu(t)] &=& -\dnu\xmu(q(t)) \pmu(t) +\nl
 &&+ \summn p
 \rep^\al_\bt(T^\mm_\nn(\dnu\xi(q(t)))) \fm\al(t) \pin\bt(t), \nl
{[}\J_X, \pnu(t)] &=&  \summn p
 \rep^\al_\bt(J^\mm_\nn(\dnu X(q(t)))) \fm\al(t) \pin\bt(t), 
\label{LJp} \\
{[}L_f,\pnu(t)] &=& -f(t)\dot\pnu(t) - \dot f(t)\pnu(t),
\nnl
{[}\Lxi, \pi_e(t)] &=& {[}\J_X, \pi_e(t)] = 0, \nl
{[}L_f, \pi_e(t)] &=& -f(t)\dot\pi_e(t).
\label{Le} 
\ees
Actually, this transformation law is more general than what follows from
(\ref{gen}), because we have included an extra term proportional to
the parameter $\la$ in (\ref{Lfn}), without spoiling the
representation condition. We call $\la$ the {\em causal weight},
and note that the einbein $e(t)$ is a zero-jet in the trivial
$\gloj$ representation with unit causal weight.

The observer's trajectory does not transform as a zero-jet,
but its derivative does (with causal weight one):
\bes
[\Lxi,\dot\qmu(t)] &=& \dnu\xmu(q(t))\dot\qnu(t), \nl
{[}\J_X, \dot\qmu(t)] &=& 0, \\
{[}L_f,\dot\qmu(t)] &=& -f(t)\ddot\qmu(t) - \dot f(t)\dot\qmu(t).
\eens

In view of the field-dependent terms in (\ref{LJp}), $\pmu(t)$ no 
longer transforms in a simple fashion under $diff(N)$. However,
\be
P_\mu(t) = \pmu(t) + \sum_{\mm} \phi_{\al,\mm+\mmu}(t)\pim\al(t),
\ee
has a simple transformation law in the infinite jet space $J^\infty\PP$.
This formula suggests that we define the total derivative as
\be
\dd_\mu f = \int dt\ [P_\mu(t), f], \qquad
\forall f \in \CF(J^\infty\QQ).
\label{totder}
\ee
The name is motivated by the following formulas:
\be
\dd_\mu q^\nu(t) = \dlt^\nu_\mu, \qquad
\dd_\mu \fn\al(t) = \phi_{\al,\nn+\mmu}(t).
\ee
In the finite jet case, the total derivative is a map 
$\dd_\mu: \CF(J^p\QQ) \longrightarrow \CF(J^{p+1}\QQ)$.

\subsection{ Fock space and normal ordering }
All functions over $S^1$ can be expanded in a Fourier series, e.g.,
\bes
\fm\al(t) &=& \sum_{r=-\infty}^\infty 
 \hphi_{\al,\mm}(r) \e^{-irt} \equiv
\fm\al^<(t) + \fm\al^\geq(t), \nle
\pim\al(t) &=& \sum_{r=-\infty}^\infty
 \hpi^{\al,\mm}(r) \e^{-irt} \equiv
\pim\al_\leq(t) + \pim\al_>(t), \nl
\eens
where the sums in 
$(\fm\al^<(t), \fm\al^\geq(t), \pim\al_\leq(t), \pim\al_>(t))$
are taken over (negative, non-negative, non-positive, positive)
frequency modes only. Similarly, $\qmu(t)$, $\pnu(t)$, $e(t)$
and $\pi_e(t)$ are divided into positive and negative frequency
modes.

Quantization amounts to replacing the Poisson brackets (\ref{Poisson})
by graded commutators;
the Fock space $J^p\FF$ is the universal envelopping
algebra of (\ref{Poisson}) modulo relations
\be
\qmu_<(t)\ket0 = \pmu^\leq(t)\ket0 = \fm\al^<(t)\ket0 = 
\pim\al_\leq(t)\ket0 = e_<(t)\ket0 = \pi_e^\leq(t) \ket0
 = 0.
\label{ket0}
\ee
The dual Fock space $J^p\FF'$ is built from a dual vacuum $\bra0$, 
annihilated by the remaining operators.
\be
\bra0\qmu_\geq(t) = \bra0\pmu^>(t) = \bra0\fm\al^\geq(t) =
\bra0\pim\al_> = \bra0 e_\geq(t) = \bra0\pi_e^>(t).
\label{bra0}
\ee
Eqs. (\ref{ket0}) and (\ref{bra0}) together imply that the
vacuum expectation value
$\bra0\fm\al(t)\ket0 = 0$ for every $\fm\al(t)$. This is not 
consistent with the condition that the metric and einbein have
inverses. Therefore we define
$g_{\mu\nu,\mm}(t) = \eta_{\mu\nu}\dlt_\mm + h_{\mu\nu,\mm}(t)$
and $e(t) = 1 + e'(t)$, where $\eta_{\mu\nu}$ is the flat Minkowski
metric, and rather demand that $h_{\mu\nu,\mm}(t)$ and $e'(t)$ 
satisfy (\ref{ket0}) and (\ref{bra0}). Note that this decomposition is
quite general; $h_{\mu\nu,\mm}(t)$ is not required to be small, only
to have vanishing vacuum expectation value.
Moreover, the geodesic equation in vacuum reads $\ddot\qmu(t) = 0$,
so $\qmu(t)$ may contain a linear part with non-zero vacuum expectation
value.

The Fock spaces $J^p\FF$ and $J^p\FF'$ are {\em not} isomorphic.

Normal ordering is necessary to remove infinites and to obtain a well 
defined action on the Fock space. For every 
$F(q, e, \phi) \in \CF(J^p\QQ)$ 
(independent of all canonical momenta), denote
\bes
\no{F(q, e, \phi)\pmu(t)} 
 &=& F(q, e, \phi)\pmu^\leq(t) + \pmu^>(t)F(q, e, \phi), 
\nle
\no{ \fm\al(t)\pin\bt(t) } &=&
\fm\al(t) \pin\bt_\leq(t) + (-)^{\al\bt} \pin\bt_>(t)\fm\al(t).
\eens

\subsection{ Some definitions }
Before describing the $DGRO(N)$ action on $J^p\FF$, some more 
preparation is needed. Let $A = (A^\al_\bt)$ be a matrix acting on $V$. 
Its supertrace is $\str A = (-)^\al A^\al_\al 
= \sum_{\al\ \bosonic} A^\al_\al 
- \sum_{\al\ \fermionic} A^\al_\al$.
For every $\gloj$ representation $\rep$ acting on $V$, define the 
numbers  $\sd(\rep)$ (super dimension), $k_0(\rep)$, $k_1(\rep)$, 
$k_2(\rep)$, $y(\rep)$, $z(\rep)$, and $k_z(\rep)$ by
\bes
\str(I) &=& \sd(\rep), \nl
\str(T^\mu_\nu) &=& k_0(\rep) \dlt^\mu_\nu, \nl
\str(T^\mu_\nu T^\si_\tau) &=&
 k_1(\rep) \dlt^\mu_\tau \dlt^\si_\nu 
 + k_2(\rep) \dlt^\mu_\nu \dlt^\si_\tau, 
\label{pardef} \\
\str(J^a) &=& z(\rep)\dlt^a, \nl
\str(J^aJ^b) &=& y(\rep) \dlt^{ab} \nl
\str(J^a T^\mu_\nu) &=& k_z(\rep) \dlt^a \dlt^\mu_\nu.
\eens
Since $gl(N) \cong sl(N)\oplus gl(1)$, its generators can be written as
\be
T^\mu_\nu = S^\mu_\nu + \omega \dlt^\mu_\nu I,
\ee
where $S^\mu_\nu = T^\mu_\nu - (1/N)\dlt^\mu_\nu T^\rho_\rho$ are the
generators of $sl(N)$, $S^\mu_\mu = 0$, and $\str(S^\mu_\nu) = 0$. 
For our purposes, $sl(N) \cong A_{N-1} = su(N)$, and 
$\dimm sl(N) = N^2-1$.
The scalar $\omega$, which labels the $gl(1)$
irreps, is related to the $gl(N)$ weight $\ka$: $\omega = -\ka+p-q$,
where $p$ ($q$) denotes the number of upper (lower) tensor indices.
Every $\gloj$ representation can be written as
$\rep = \sum_{i\in I} R_i\oplus\omega_i\oplus M_i$, 
where $R_i$, $\omega_i$ and $M_i$ denote irreducible $sl(N)$, $gl(1)$ 
and $\oj$ representations and $I$ is some index set. 
To each irrep we assign a Grassmann parity factor $(-)^i$.
The parameters in (\ref{pardef}) can then be written
\bes
\sd(\rep) &=& \sum_{i\in I} (-)^i \dimm R_i \dimm M_i, \nl
k_0(\rep) &=& \sum_{i\in I} (-)^i k_0(R_i,\omega_i)\dimm M_i, \nl
k_1(\rep) &=& \sum_{i\in I} (-)^i k_1(R_i) \dimm M_i, \nl
k_2(\rep) &=& \sum_{i\in I} (-)^i k_2(R_i, \omega_i)\dimm M_i, 
\label{pars} \\
y(\rep) &=& \sum_{i\in I} (-)^i \dimm R_i\, y(M_i), \nl
z(\rep) &=& \sum_{i\in I} (-)^i \dimm R_i\, z(M_i), \nl 
k_z(\rep) &=& \sum_{i\in I} (-)^i k_0(R_i,\omega_i)\, z(M_i).
\eens
For $R\oplus\omega$ a $gl(N)$ irrep, 
\bes
k_0(R, \omega) &=& \omega \dimm R, \nl
k_1(R) &=& {2x_R\/\dimm R}, \\
k_2(R, \omega) &=& \omega^2 \dimm R - {2x_R\/N\dimm R},
\eens
where $x_R$ is the Dynkin index (a positive integer) of the 
representation $R$; it is related to the value of the quadratic 
Casimir as $Q_R = 2x_R(N^2-1)/\dimm R$. For $\oj$ semisimple, 
\be
z(M)=0, \qquad y(M) = {2\dimm \oj\/\dimm M} x_M,
\ee
where $x_M$ is the Dynkin index of the representation $M$
\cite{FMS96, GO86}.

The calculation of the abelian charges will use the following
results \cite{Lar98}.
\bes
&&\summ{p} \dlt^\mm_\mm\, \str(I) = \Np{} \sd(\rep), \nl
&&\summ{p} \str(T^\mm_\mm(\xi))
= \dmu\xmu \Big( \Np{} k_0(\rep) - \Np{-1} \sd(\rep) \Big), \nl
&&\sum_{|\mm|,|\nn|\leq p} \str( T^\mm_\nn(\xi) T^\nn_\mm(\eta) )
= \dnu\xmu\dmu\ynu \Big( \Np{} k_1(\rep)
+ \Np{-1} \sd(\rep) \Big) +\nl
\bl+ \dmu\xmu\dnu\ynu \Big( \Np{} k_2(\rep) +{N+1\/N}\Np{-2}\sd(\rep)
- 2\Np{-1} k_0(\rep) \Big), \nl
&&\summ{p}  \str(J^\mm_\mm(X))
= X_a z(\rep)\dlt^a \Np{}, \\
&&\sum_{|\mm|,|\nn|\leq p} \str( J^\mm_\nn(X) J^\nn_\mm(Y))
= y(\rep) \Np{} X_aY_b\dlt^{ab}, \nl
&&\sum_{|\mm|,|\nn|\leq p} \str(T^\mm_\nn(\xi) J^\nn_\mm(X)) = \nl
\bl = \dmu\xmu X_a \dlt^a \Big( \Np{}k_z(\rep) - \Np{-1} z(\rep) \Big).
\eens
Compared to \cite{Lar98}, a minus sign for fermions has been absorbed
into the definition (\ref{pars}).

\subsection{ $DGRO(N)$ action on $J^p\FF$ }
The main result of \cite{Lar98} is the explicit description of the
DGRO algebra action on $J^p\FF$. It follows from theorems 5.1 and
6.2 in that paper that the following operators 
provide a realization of the $DGRO(N)$.
\bes
\Lxi &=& \int dt\ \no{\xmu(q(t)) \pmu(t)} - \summn{p}
  \rep^\al_\bt(T^\mm_\nn(\xi(q(t)))) \no{ \fm\al(t)\pin\bt(t)  } +\nl
&&+ {u_1\/2\pi i} \int dt\ \dmu\xmu(q(t)), \nl
\J_X &=& -\summn{p}\int dt\ 
  \rep^\al_\bt(J^\mm_\nn(X(q(t)))) \no{ \fm\al(t)\pin\bt(t) } +\nl
&&+ {u_2\/2\pi i} \dlt^a \int dt\ X_a(q(t)), 
\label{LJL} \\
L_f &=& \int dt\  f(t)( - \no{\dot\qmu(t)\pmu(t)}
 - \no{ \dot e(t)\pi_e(t) }
 -  \summ{p} \no{ \dot\fm\al(t)\pim\al(t)} ) -\nl
&&- \dot f(t) ( \no{ e(t)\pi_e(t) }
  + \la \summ{p} \no{ \fm\al(t)\pim\al(t)} ) + {u_3\/2\pi i},
\eens
Compared to \cite{Lar98}, the two terms proportional to $u_1$ and $u_2$ 
have been added, to fix normalization of the trivial 
cocycles in (\ref{DRO}) and (\ref{DGRO}). 
\bes
u_1 &=& -\la 
 \Big( \Np{}k_0(\rep) - \Np{-1} \sd(\rep) \Big), \nl
u_2 &=& - \la\, z(\rep) \Np{}, \\
u_3 &=& \half (\la - \la^2) \sd(\rep) \Np{},
\eens
which define funtions $u_j(p,N;\rep,\la)$, $j=1,2,3$.
The presence of these terms, as well as normal ordering, modifies the
transformation law (\ref{LJp}) for $\pnu(t)$.
The abelian charges,
\be
c_j = \cq_j(N) + \ce_j(1) + \cf_j(p,N;\rep,0),
\label{cj}
\ee
are given in terms of functions 
$\cq_j(N)$ (contribution from the observer's trajectory),
$\ce_j(\la)$ (contribution from the einbein), and
$\cf_j(p,N;\rep,\la)$ (contribution from the fields).
\bes
\cq_1(N) &=& 1, \qquad \cq_2(N) = 0, \nl
\cq_3(N) &=& 1, \qquad \cq_4(N) = 2N, 
\label{cq}\\
\cq_5(N) &=& \cq_6(N) = \cq_7(N) = 0, 
\nnl
\ce_4(\la) &=& 2(1 - 6\la+6\la^2), \qquad
\ce_j(\la) = 0 \hbox{ otherwise,} \\
\nl
\cf_1(p,N;\rep,\la) &=& \Np{}k_1(\rep) + \Np{-1} \sd(\rep), \nl
\cf_2(p,N;\rep,\la) &=& \Np{}k_2(\rep) +\nl
&+&{N+1\/N}\Np{-2} \sd(\rep) - 2\Np{-1} k_0(\rep), \nl
\cf_3(p,N;\rep,\la) &=& (2\la-1)
\Big( \Np{}k_0(\rep) - \Np{-1} \sd(\rep) \Big), \nl
\cf_4(p,N;\rep,\la) &=& 2(1 - 6\la+6\la^2) \Np{} \sd(\rep), 
\label{cf} \\
\cf_5(p,N;\rep,\la) &=& -\Np{}y(\rep), \nl
\cf_6(p,N;\rep,\la) &=& (2\la-1) \Np{} z(\rep), \nl
\cf_7(p,N;\rep,\la) &=& \Np{-1}z(\rep) - \Np{}k_z(\rep).
\eens
Define
\be
\tcf_j(p,N;\rep,\la) = \cf_j(p,N+1;\rep,\la) - \cf_j(p-1,N+1;\rep,\la).
\label{tcf}
\ee
For future reference we record the formulas
\bes
{N+p\choose p} - {N+p-1\choose p-1} &=& {N-1+p\choose p}, \nl
{N+p\choose p-1} - {N+p-1\choose p-2} &=& {N-1+p\choose p-1}, 
\label{Np1} \\
{N+p\choose p-2} - {N+p-1\choose p-3} &=& {N-1+p\choose p-2},
\eens
which imply that
$\tcf_j(p,N;\rep,\la) = \cf_j(p,N;\rep,\la)$,
if $j\neq2$.
An analogous formula holds for $u_j$:
\be
u_j(p,N;\rep,\la) - u_j(p-1,N;\rep,\la) = u_j(p,N-1;\rep,\la).
\label{ured}
\ee

\section{Constraints in jet space}

\subsection{ State cohomology }
The Fock spaces described in the previous sections are good quantum
theories in the sense that they carry well-defined representations of 
$DGRO(N,\oj)$, with a non-trivial abelian extension. However, neither
are they reducible, nor is dynamics (EL equations) taken into
account. The strategy for constructing smaller modules
is to take the KT generator $\QKT$ from section \ref{sec:class},
expand in a Taylor series in $(x-q(t))$, and discard all terms
involving derivatives of order higher than $p$. Since the KT
generator is invariant, the state cohomology defines a 
$DGRO(N,\oj)$ module.

Define a physical state $\ket{phys} \in J^p\FF$ as a state that is
annihilated by the KT generator, $\QKT\ket{phys} = 0$.
The state cohomology $\Hstate \equiv \Hstate(\QKT, J^p\FF)$
is the space of physical states modulo relations 
$\ket{phys} \sim \ket{phys} + \QKT\ket{}$,
i.e. the cohomology of the complex 
$\Omega^\bullet_{state} = \sum_{g=-\infty}^\infty  \Omega^g_{state}$,
where 
$\Omega^g_{state} = \{ \Psi_g\ket0 : \Psi_g\in J^p\PP, \gh\,\Psi_g = g\}$.
For $\Hstate$ to be well defined, $\QKT$ must be normal 
ordered. However, $\QKT$ is always bilinear in
commuting variables, so normal ordering has no effect. This is the
crucial reason to prefer the KT generator over the BRST one.

The dual state cohomology $\Hstate(\QKT, J^p\FF')$ is the space of
dual physical states $\bra{phys'} \in J^p\FF'$, satisfying
\be
\bra{phys'}\QKT = 0, \qquad  
\bra{phys'} \sim \bra{phys'} + \bra{}\QKT.
\ee
The two cohomologies are 
not isomorphic, since the underlying spaces $J^p\FF$ and $J^p\FF'$
are not so. However, two states of fixed ghosts number, 
\be
\ket{g,phys} = \Psi_g\ket0, \qquad
\bra{phys',g} = \bra0\Psi_{g}',
\ee
where $\gh\,\Psi_g = \gh\,\Psi_{g}' = g$, satisfy orthogonality
conditions of the form
\be
\langle{phys',g'}\,|\,{g,phys}\rangle \propto \dlt_{g+g'}.
\label{ortho}
\ee
In the remainder of this section we focus on the ket state cohomology
$\Hstate$ only.

A physical operator $A_{phys}$ satisfies
\be
[\QKT, A_{phys}] = 0, \qquad A_{phys} \sim A_{phys} + [\QKT, C].
\label{physop}
\ee
Only physical operators act in a well defined manner on 
$\Hstate$, and hence we must demand that
that all $DGRO(N,\oj)$ generators are physical operators. Note that
normal ordering does not affect the conditions (\ref{physop}) if
$A_{phys}$ is at most linear in the momenta, because normal ordering
has no effect on $\QKT$.

The momentum number $\mom$ is no longer well defined, since operators
of non-zero momentum number are created out of the vacuum. In the
simplest case where there are neither fields nor einbein,
\be
L_{\exp(imt)}\ket0 = \int dt\  \e^{imt}\pmu^>(t)\dot\qmu(t)\ket0,
\ee
which is non-empty for positive $m$. Hence $\Omega^g_{state}$ can not
be further decomposed into states of fixed momentum number, and all
cohomology groups $H^g_{state}$ are in general non-zero.
However, physical states of non-zero ghost number decouple due to the
orthogonality relation (\ref{ortho}), and each cohomology group
is a well-defined $DGRO(N,\oj)$ module.

\subsection{ Longitudinal constraint }
The Taylor coefficents depend on the parameter $t$ although the field
itself does not, because the expansion point $\qmu(t)$ does.
On the other hand, the RHS of (\ref{gen}) actually defines a function
$\fa(x,t)$ of two variables. To resolve this paradox we must impose
the condition ${\d \fa(x,t)/\d t} = 0$, which is equivalent to
\be
\DD_{\al,\mm}(t) \equiv
\dot\fm\al(t) - \dot\qmu(t)\phi_{\al,m+\mmu}(t) \approx 0.
\label{DD1}
\ee
Introduce an antifield $\bt_{\al,\mm}(t)$, with momentum 
$\gm^{\al,\mm}(t)$, of opposite Grassman parity, and subject to
\be
[\gm^{\al,\mm}(s), \bt_{\bt,\nn}(t)] =
 \dlt^\al_\bt \dlt^\mm_\nn \dlt(s-t).
\ee
$\fm\al(t)$ is defined for $|\mm|\leq p$, but $\DD_{\al,\mm}(t)$
(and thus $\bt_{\al,\mm}(t)$ and $\gm^{\al,\mm}(t)$)
are only defined for $|\mm|\leq p-1$, due to the appearence of
$\phi_{\al,m+\mmu}(t)$ in (\ref{DD1}). Moreover, the $t$ derivative
makes the causal weight equal one rather than zero.
$DGRO(N,\oj)$ thus acts on the antifields as
\bes
{[}\Lxi,\bt_{\al,\nn}(t)] &=& 
 -\summ{|\nn|} \rep^\bt_\al(T^\mm_\nn(\xi(q(t)))) \bt_{\bt,\mm}(t), \nl
{[}\J_X,\bt_{\al,\nn}(t)] &=& 
 -\summ{|\nn|} \rep^\bt_\al(J^\mm_\nn(X(q(t)))) \bt_{\bt,\mm}(t), 
\label{LJbt} \\
{[}L_f,\bt_{\al,\nn}(t)] &=& 
 -f(t)\dot\bt_{\al,\nn}(t) - \dot f(t)\bt_{\al,\nn}(t), 
\nnl
{[}\Lxi,\gm^{\al,\mm}(t)] &=& 
 \summn{p-1} \gm^{\bt,\nn}(t) \rep^\al_\bt(T^\mm_\nn(\xi(q(t)))), \nl
{[}\J_X,\gm^{\al,\mm}(t)] &=& 
 \summn{p-1} \gm^{\bt,\nn}(t) \rep^\al_\bt(J^\mm_\nn(X(q(t)))), 
\label{LJgm} \\
{[}L_f,\gm^{\al,\mm}(t)] &=& -f(t)\dot\gm^{\al,\mm}(t).
\eens
Equivalently, the contributions to the $DGRO(N)$ generators are
\bes
\Lxi^{(\DD)} &=& - \summn{p-1} \int dt\
  \rep^\al_\bt(T^\mm_\nn(\xi(q(t)))) 
  \no{ \bt_{\al,\mm}(t)\gm^{\bt,\nn}(t)}, \nl
\J_X^{(\DD)} &=& - \summn{p-1} \int dt\
   \rep^\al_\bt(J^\mm_\nn(X(q(t))))
  \no{ \bt_{\al,\mm}(t)\gm^{\bt,\nn}(t)}, 
\label{LJD} \\
L_f^{(\DD)} &=&  \summn{p-1} \int dt\
 f(t)\no{\bt_{\al,\mm}(t)\dot\gm^{\al,\mm}(t)},
\eens
apart from cohomologically trivial terms.
The contribution to the KT generator is
\be
\QKT^{(\DD)} = \summ{p-1} \int dt\ \DD_{\al,\mm}(t) \gm^{\al,\mm}(t).
\ee
Note that no normal ordering is necessary here.
The antifield contribution to the abelian charges is readily 
computed in terms of the functions (\ref{cf}). $\bb_{\al,\mm}(t)$
has causal weight one, is defined for $|\mm|\leq p-1$, and has
opposite Grassmann parity, so its contribution counts negative.
Hence the abelian charges are
\be
c_j = \cq_j(N) + \ce_j(1) + \cf_j(p,N;\rep,0) - \cf_j(p-1,N;\rep,1).
\ee

Instead of (\ref{DD1}), we can consider the alternative constraint
\be
\DD_{\al,\mm}(t) = e^{-1}(t)(\dot\fm\al(t) 
 - \dot\qmu(t)\phi_{\al,\mm+\mmu}(t)) \approx 0,
\label{DD2}
\ee
where $e^{-1}(t)$ is the inverse of the einbein
$e(t)$, corresponding to the equally true fact 
$e^{-1}(t){\d \fa(x,t)/\d t} = 0$. The expression in (\ref{DD2})
has causal weight zero. This change induces the following
modifications in (\ref{LJbt}), (\ref{LJgm}) and (\ref{LJD}):
\bes
{[}L_f,\bt_{\al,\nn}(t)] &=& 
 -f(t)\dot\bt_{\al,\nn}(t), \nl
{[}L_f,\gm^{\al,\mm}(t)] &=& 
 -f(t)\dot\gm^{\al,\mm}(t) - \dot f(t)\gm^{\al,\mm}(t) \\
L_f^{(\DD)} &=& - \summn{p-1} \int dt\
 f(t)\no{\dot\bt_{\al,\mm}(t)\gm^{\al,\mm}(t)}.
\eens
Using (\ref{tcf}), the abelian charges now become
\be
c_j = \cq_j(N) + \ce_j(1) + \tcf_j(p,N-1;\rep,0).
\label{cj2}
\ee
This result has a natural
interpretation: when (\ref{DD2}) is taken into account, only the
transverse modes $\fm\al(t) - \dot\qmu(t)\phi_{\al,\mm+\mmu}(t)$ 
contribute.
They are equal in number to $\fm\al(t)$ with $m_0 = 0$, i.e.
the dimension is effectively reduced from $N$ to $N-1$.

Thus the cohomologies of the equivalent constraints 
(\ref{DD1}) and (\ref{DD2}) give rise to inequivalent $DGRO(N,\oj)$ 
modules.
A deeper understanding of this disturbing fact is lacking. 
For definiteness, only the second form (\ref{DD2}) is considered 
henceforth, so the abelian charges are given by (\ref{cj2}).

\subsection{ Euler-Lagrange constraint }
Since the EL constraint (\ref{EL}) is a local functional, it can be
expanded in a Taylor series,
\be
\Ea(x) = \sum_{|\mm|\geq0} {1\/\mm!} \Eam(t)(x-q(t))^\mm.
\ee
Assume that the EL equations are of order $o_\al$, which typically
has the values $o_\al = (2, 1, 0)$ for $\fa$ a 
(bosonic, fermionic, auxiliary) degree of 
freedom. Then the Taylor coefficients $\Eam(t)$ is a function 
of $\fn\bt(t)$ for $|\nn| \leq |\mm| + o_\al$, which is well defined on
$J^p\QQ$ provided that $|\mm| \leq p - o_\al$.
Thus, $\Eam(t)$ transforms as a $(p-o_\al)$-jet with an
upper $V$ index and a lower multi-index, and with causal weight zero.
The EL constraint now takes the form 
\be
\Eam(t) \approx 0, \qquad
\forall\, |\mm| \leq p - o_\al.
\label{ELm}
\ee
The antifield $(p-o_\al)$-jet $\fsm\al(t)$ is introduced to kill the EL
equation (\ref{ELm}) in cohomology; it can be considered as the Taylor
coefficients of $\fsa(x)$ up to order $p-o_\al$. The corresponding
momentum $\psm\al(t)$ is defined by relations
\be
[\psm\al(s), \fsn\bt(t)] = \dlt_\al^\bt \dlt^\mm_\nn \dlt(s-t).
\ee
The contributions to the $DGRO(N)$ generators are
\bes
\Lxi^{(\EE)} &=& - \summn{p-o_\al} \int dt\ 
  \rep^\al_\bt(T^\mm_\nn(\xi(q(t)))) 
  \no{ \fsm\bt(t) \psn\al(t) }, \nl
\J_X^{(\EE)} &=& - \summn{p-o_\al} \int dt\ 
   \rep^\al_\bt(J^\mm_\nn(X(q(t))))
  \no{ \fsm\bt(t) \psn\al(t) },
\label{LJE} \\
L_f^{(\EE)} &=& - \summ{p-o_\al}\int dt\ 
 f(t)\no{\dot\fsm\al(t)\psm\al(t)},
\eens
apart from cohomologically trivial terms.

To make the notation completely clear: the representation $\rep$ is
a direct sum, 
$\rep = \rep_{\bosonic}\oplus\rep_{\fermionic}\oplus\rep_{\auxiliary}$,
so $\Lxi^{(\EE)} = \LL_{\xi,\bosonic}^{(\EE)} 
+ \LL_{\xi,\fermionic}^{(\EE)} + \LL_{\xi,\auxiliary}^{(\EE)}$,
and the sums in (\ref{LJE}) runs over
$|\nn|\leq p-2$, $p-1$, and $p$, respectively.

Since $\fsm\al(t)$ carries an upper $V$ index, it
transforms in the dual $\gloj$ representation $\rep^\dagger$, just
like $\pim\al(t)$.
The contribution to the KT generator is
\be
\QKT^{(\EE)} = \summ{p-o_\al} \int dt\ \Eam(t) \psm\al(t),
\ee
which adds $-\cf(p-o_\al, N; \rep^\dagger, 0)$ to the abelian
charges.

However, the constraints (\ref{DD2}) and (\ref{ELm}) are not
independent, because
\be
\BB^\al_{,\mm}(t) = 
e^{-1}(t)(\dot\Eam(t) - \dot\qmu(t)\Ea_{,\mm+\mmu}(t) )
\approx 0,
\ee
for every $\mm$ such that $|\mm|\leq p-o_\al-1$. 
New antifields must therefore be introduced to eliminate the unwanted 
cohomology; call these $\bsm\al(t)$ and their momenta
$\csm\al(t)$. The contribution to the KT generator is
\be
\QKT^{(\BB)} = \summ{p-o_\al-1} \int dt\ 
 \BB^\al_{,\mm}(t) \csm\al(t).
\ee
Similarly, there are contributions to $\Lxi$, $\J_X$ and $L_f$,
analogous to (\ref{LJE}), but the sum only runs up to $p-o_\al-1$.
These antifields have opposite Grassmann parity from $\fsm\al(t)$
and $\bt_{\al,\mm}(t)$, and thus the same parity as the original
field $\fm\al(t)$. As the notation suggests, we can view them as the 
antifields of $\bt_{\al,\mm}(t)$. Their addition to the abelian charges 
is $\cf(p-o_\al-1, N; \rep^\dagger, 0)$.
In view of (\ref{tcf}), 
the net contribution to the abelian charges from the antifields is thus
\bes
&&-\cf(p-o_\al, N; \rep^\dagger, 0) 
+\cf(p-o_\al-1, N; \rep^\dagger, 0) \nle
&&= -\tcf(p-o_\al, N-1; \rep^\dagger, 0).
\eens

\subsection{Geodesic constraint}
The geodesic and einbein constraints are modified as follows
in the passage to jet space. 
In (\ref{geo}), replace $g_{\mu\nu}(q(t))$ and 
$\Gamma^\nu_{\si\tau}(q(t))$ by the zero-jets $g_{\mu\nu}(t)$ and 
$\Gamma^\nu_{\si\tau}(t)$, and in the definition of the latter 
(\ref{LC}), replace $\drho g_{\mu\nu}(q(t))$ by $g_{\mu\nu,\rho}(t)$.
The trajectory antifield thus contributes
$-\cq_j(N)$ to the abelian charges, which cancels the contribution from
the observer's trajectory. The einbein antifield has causal weight
zero, but since $\ce_j(1) - \ce_j(0) = 0$, (both terms equal
$2\dlt_{j,4}$), the net result from the einbein is zero.

\subsection{ Noether identities}
Finally, we must consider the Noether symmetries. In (\ref{Noether}),
we expand $\Ea(y)$ in a Taylor series. By invariance, it is now clear 
that there must exist some functions $r^{a,\nn}_\al(x,t)$, such that
\be
r^a(x) = \sum_\nn \int dt\ r^{a,\nn}_\al(x,t)\Ea_{,\nn}(t).
\ee
Hence the corresponding jet, obtained by a Taylor expansion 
in $x$, has the form
\be
r^a_{,\mm}(s) = \sumn{|\mm|} \int dt\ 
 r^{a,\nn}_{\al,\mm}(s,t)\Ea_{,\nn}(t).
\ee
This formula defines functions $r^{a,\nn}_{\al,\mm}(s,t)$, whose
transformation properties is clear from their index structure.
The Noether identities now turn into operator identities 
$r^a_{,\mm}(s)\equiv 0$,
valid for all $\mm$ of sufficiently low order, say $|\mm|\leq p-o_a$,
where $o_a$ is the order of the original Noether identity 
(\ref{Noether}). Thus for every physical state $\ket{phys}$, 
\be
\sumn{|\mm|} \int dt\ 
 (-)^\al r^{a,\nn}_{\al,\mm}(s,t)\fsn\al(t) \ket{phys}
\label{jetclosed}
\ee
is also physical. To eliminate this unwanted cohomology, we must
introduce additional (bosonic) Noether antifields to make
(\ref{jetclosed}) exact.
Denote these jets $\bb^a_{,\mm}(t)$ and their momenta
$\cc^{,\mm}_a(t)$.
To (\ref{bc}) and (\ref{Q2}) correspond
\bes
&&[\cc^{,\mm}_a(s), \bb^b_{,\nn}(t)] 
= \dlt^b_a \dlt^{,\mm}_{,\nn}\dlt(s-t), \\
&&\QKT^{(2)} = \sumn{|\mm|\leq p-o_a} \iint dsdt\ (-)^\al 
r^{a,\nn}_{\al,\mm}(s,t)\fsn\al(t)\cc^{,\mm}_a(s).
\eens
Now the state in (\ref{jetclosed}) can be written as
$\QKT \bb^a_{,\mm}(t)\ket{phys}$, and thus it no longer contributes
to the cohomology.

Now specialize to $DGRO(N,\oj)$. 
To each type of symmetry, we assign bosonic antifields 
and momenta according to the following table.
\[
\begin{array}{ccccc}
\hbox{symmetry} & \hbox{antifield} & \hbox{af. momentum} 
 &\hbox{antifield jet} & \hbox{af. momentum jet} \\
diff(N) & \bb_\mu(x) & \cc^\mu(x) & \bb_{\mu,\mm}(t) 
 & \cc^{\mu,\mm}(t) \\
diff(1) & \bb(t) & \cc(t) & \bb(t) & \cc(t) \\
map(N,\oj) & \bb^a(x) & \cc_a(x) & \bb^a_{,\mm}(t) & \cc_a^{,\mm}(t)
\end{array}
\]
The Noether identities for $diff(N)$ 
and $map(N,\oj)$ are both of order three, because the dominating
terms are $\dnu G^{\mu\nu}(x) \equiv 0$ and 
$\dmu \dnu F^{\mu\nu}(x) \equiv 0$, respectively. Therefore,
$\bb_{\mu,\mm}(t)$ and $\bb^a_{,\mm}(t)$ are both defined for
$|\mm|\leq p-3$, while the reparametrization antifield is not
affected by the passage to jet space.

Define the following fields:
\bes
\SS_\mu(x) &=& \int \dNy (-)^\al[\LL_\mu(x), \fa(y)] \fsa(y), \nl
\TT_\mu(x) &=& \int dt\ [\LL_\mu(x), \qnu(t)] q^*_\nu(t), \\
\WW^a(x) &=& \int \dNy (-)^\al [\J^a(x), \fa(y)] \fsa(y).
\eens
Note that $\TT_\mu(x) \propto \dlt^N(x-q(t))$.
If $\SS_{\mu,\mm}(t)$, $\TT_{\mu,\mm}(t)$, and $\WW^a_{,\mm}(t)$
denote the corresponding jet space trajectories,
the KT generator contributions (\ref{QNoet}) become
\bes
\QKT^{(\diff)} &=& \summ{p-3}
 \int dt\ (\SS_{\mu,\mm}(t) + \TT_{\mu,\mm}(t)) \cc^{\mu,\mm}(t), 
\nl
\QKT^{(\rrep)} &=&  
\iint dsdt\ ( [L(s), \qmu(t)] q^*_\mu(t) + [L(s), e(t)] e^*(t)) \cc(s), 
\\
\QKT^{(\gauge)} &=& \summ{p-3} 
\int dt\ \WW^a_{,\mm}(t) \cc_a^{,\mm}(t),
\eens
However, the Noether antifields are not all independent, because there
are further conditions analogous to (\ref{DD2}).
\bes
e^{-1}(t)( \dot \bb_{\mu,\mm}(t) - \dot\qmu(t) \bb_{\mu,\mm+\mmu}(t))
 &\approx& 0, 
\nllbnl{extra}
e^{-1}(t)( \dot \bb^a_{,\mm}(t) - \dot\qmu(t) \bb^a_{\mu,\mm+\mmu}(t) )
 &\approx& 0.
\eens
We must thus introduce further antifields to eliminate these relations
in cohomology.

It is now straightforward to write down the Noether antifield 
contributions to the DGRO algebra generators. Suffice it to say, that
they transform in the adjoint representation of $DGRO(N)$, which
corresponds to the $\gloj$ representation $\Ad$
and causal weight one. Here $\ad_\oj$ denotes the $\oj$ adjoint and 
$0_\oj$ the trivial $\oj$ representation.
This is distributed among the various           
Noether antifields according to the following table.
\[
\begin{array}{ccccc}
\hbox{symmetry} & \hbox{antifield} & gl(N)\hbox{ rep} 
&\oj\hbox{ rep} & \hbox{causal weight} \\
diff(N) & \bb_\mu(x) & (1,0;0) & 0_\oj & 0 \\
diff(1) & \bb(t) & (0,0;0) & 0_\oj & 1 \\
map(N,g) & \bb^a(x) & (0,0;0) & \ad_\oj & 0
\end{array}
\]
Moreover, antifields for the conditions (\ref{extra}) must also 
be considered.

Hence we get for the abelian charges
\bes
c_j^{(\diff)} &=& \cf_j(p-3, N; (1,0;0)\oplus 0_\oj, 0)
- \cf_j(p-4, N; (1,0;0)\oplus 0_\oj, 0) \nl
&=& \tcf_j(p-3, N-1; (1,0;0)\oplus 0_\oj, 0), \nl
c_j^{(\rrep)} &=& \ce_j(1) = 2 \dlt_{j,4}, \\
c_j^{(\gauge)} &=& \cf_j(p-3, N; (0,0;0)\oplus \ad_\oj, 0)
- \cf_j(p-4, N; (0,0;0)\oplus \ad_\oj, 0) \nl
&=& \tcf_j(p-3, N-1; (0,0;0)\oplus \ad_\oj, 0).
\eens
These three terms can be summed to yield the following total
contribution from the Noether symmetries:
\be
c_j^{(\Noether)} = 2 \dlt_{j,4} + \tcf_j(p-3, N-1; \Ad, 1).
\ee

\subsection{ Finiteness conditions }
The total KT generator is thus
\be
\QKT = \QKT^{(\DD)} + \QKT^{(\EE)} + \QKT^{(\BB)} + \QKT^{(\GG)}
+ \QKT^{(\Noether)}.
\label{Qtot}
\ee
Similarly, the $DGRO(N,\oj)$ generators have analogous antifield
contributions, in addition to the original expressions (\ref{LJL}).
Summing up the abelian charges, we find
\bes
c_j &=& \tcf_j(p,N-1;\rep,0) 
- \tcf_j(p-o_\al, N-1; \rep^\dagger, 0) +\nl
&& + 2 \dlt_{j,4} + \tcf_j(p-3, N-1; \Ad, 1).
\label{ctot}
\ees       
The coefficients $u_j$ in front of the cohomologically trivial terms
have not been discussed. However, it is clear from the above discussion
and (\ref{ured}) that they satisfy an analogous equation, i.e.
\bes
u_j &=& u_j(p,N-1;\rep,0) 
- u_j(p-o_\al, N-1; \rep^\dagger, 0) \nle
&&+ u_j(p-3, N-1; \Ad, 1) = 0,
\eens
because $u_j(p,N;\rep,\la) \propto \la = 0$.

We have thus shown that $DGRO(N,\oj)$ acts on the cohomology \break
$\Hstate(\QKT, J^p\FF)$, with $\QKT$ given by (\ref{Qtot}), and that the
abelian charges are (\ref{ctot}). It is worth noting that these
modules are manifestly well defined, since the starting point
(i.e. the unconstrained modules constructed in the previous section)
contain no infinities and normal ordering does not affect the KT
generator. 
The construction works for all finite $p$, but the limit $p\to\infty$
may not exist. This may be mathematically satisfactory, but is too
isolationistic for physics: the limit expresses the objective reality of 
events a finite distance away from the observer. A necessary condition
for this limit to exist is that the total abelian charges (\ref{ctot})
have a finite limit. 

The conditions (\ref{ctot}) have not been analyzed in great detail, 
but some observations are immediate.
Since the functions (\ref{cf}) are polynomials in
$p$, we can write
\be
c_j = a^0_j + a^1_j p + \ldots + a^{n}_j p^n,
\ee
where $a^k_j$ and $n$ depend on $N$ and $j$ but not $p$. Hence the
finiteness conditions take the form $a^k_j = 0$, for all $k$, 
$1\leq k\leq n(N,j)$. 

The trajectory and einbein contributions are independent of $p$, so
they can be ignored. To leading order in $p$, 
\be
\Np{}\approx {p^N\/N!}, \quad
\Np{-1}\approx {p^{N+1}\/(N+1)!}, \quad
\Np{-2}\approx {p^{N+2}\/(N+2)!}.
\ee
The functions in (\ref{ctot}) take the form
\be
\tcf_j(p,N;\rep,\la) \approx a^n_j(N;\rep,\la) p^{n}
\label{clim}
\ee
where
\[
\begin{array}{lllll}
j &\qquad& n(N,j) &\qquad& a^n_j(N;\rep,\la) \\ 
1 &\qquad& N+1  &\qquad& \sd(\rep)/(N+1)! \\
2 &\qquad& N+2  &\qquad& \sd(\rep)/(N+1)(N+1)! \\
3 &\qquad& N+1  &\qquad& (1-2\la)\sd(\rep)/(N+1)! \\
4 &\qquad& N  &\qquad& 2(1-6\la+6\la^2)\sd(\rep)/N! \\
5 &\qquad& N  &\qquad& y(\rep)/N! \\
6 &\qquad& N  &\qquad& z(\rep)/N! \\
7 &\qquad& N+1    &\qquad& z(\rep)/(N+1)!
\end{array}
\]
Now consider the different contributions to $c_j$. 
{F}rom the original fields we get (\ref{clim}), and from the antifields 
$-a^n_j(N;\rep^\dagger,\la) (p-o_\al)^{n}$.
Since $\sd(\rep^\dagger) = \sd(\rep)$, $y(\rep^\dagger) = y(\rep)$,
$z(\rep^\dagger) = -z(\rep)$, and
$(p-o_\al)^{n} \approx p^{n}$, the fields and antifields
cancel to leading order in $p$. Cancellation is even
exact for auxiliary fields, but for ordinary fields, the Cauchy data
survive in cohomology and give rise to subleading terms. 
Hence the abelian charges are completely dominated for large $p$
by the Noether antifields, which transform in
the representation $\rep = \Ad$.
The factors $a^n_j$ are typically proportional to
$\sd(\rep) = \dimm  \rep_{\bosonic} - \dimm  \rep_{\fermionic}$, 
which is positive since we have assumed that there are no fermionic
Noether symmetries.

This is a significant problem, because it means that the abelian charges
always diverge. Some possible cures are:
\begin{enumerate}
\item
Introduce fermionic Noether symmetries, i.e. supersymmetry. This is
covered in our formalism apart from some additional signs appearing 
e.g. in (\ref{Q2}) and (\ref{JX}).
\item
Introduce fermionic fields without dynamics, so that no antifields
cancel their leading behaviour.
\item
Dismiss the Noether antifields altogether. Then there are non-zero
states of the form (\ref{jetclosed}), so the connection to
the classical theory in section \ref{sec:class} is looser, but 
the DGRO algebra still acts on the cohomology groups.
\item
Only consider finite $p$. To each classical action then corresponds 
the family of well-defined lowest-energy modules 
$H^g_{state}(\QKT, J^p\FF)$, but the limit $p\to\infty$ is ill defined.
\end{enumerate}
None of these options is satisfactory, but if we 
nevertheless opt for the third possibility, the leading terms cancel
between the fields and antifields, whereas the subleading terms
are of the form 
\be
c_j \propto o_\al \sd(\rep) p^{n-1}, \qquad
o_\al \sd(\rep) \equiv
 2\dimm  \rep_{\bosonic} - \dimm  \rep_{\fermionic} = 0.
\label{ccond}
\ee
Thus, the $p\to\infty$ limit can only exist if there are twice as many
bosonic non-auxiliary fields than fermionic ones. 
Lower-order terms imply further restrictions on the field content.
Since the number of conditions grows with $N$,
there is probably an upper critical dimension above which
no solutions exist.
However, one should not take this result too seriously, since the
Noether antifields were discarded.

\section{ Discussion }

In this paper a large class of projective lowest-energy representations 
of the Noether symmetries in physics has been constructed.
It can be considered as a novel approach to quantization, applicable 
to all systems including gravity (at finite $p$), although the limit
$p\to\infty$ is problematic. 
It should be stressed that this approach is very conservative.
No unobserved physics, such as extra dimensions,
higher-dimensional objects, Planck-scale discreteness, or supersymmetry
(at least not for finite $p$), needs to be assumed.
Rather, I start from classical physics as formulated in section 
\ref{sec:class}, expand all fields in a Taylor series around the
observer's present position,
replace Poisson brackets by commutators, and represent the 
resulting Heisenberg algebra on a unique Fock space. 

A unique feature is that the Noether symmetries
have consistent quantum representations,
i.e. well-defined modules of non-split, abelian, Lie algebra
extensions of the classical symmetry algebra. To my knowledge, this
issue has never been addressed before, which is not surprising
since the first interesting projective $diff(N)$ modules were 
only discovered in 1994 \cite{ERM94}. In standard canonical quantization 
of gravity, the constraints do not even classically reproduce $diff(N)$,
but only the so-called ``Dirac algebra''
(\cite{Ish91}, page 169).

Another point is the central role played by the observer. True, she
is important (albeit in different ways) in both quantum mechanics
and general relativity, but she does not enter directly into the 
core (Schr\"odinger and Einstein) equations. Here, the passage to jet
space introduces the observer directly into the core formalism.

The DGRO algebra contains non-split abelian extensions, which can be
viewed as quantum anomalies. Although anomalies often are though of
as harmful, I believe they are necessary and quite useful. In
particular, it is sometimes claimed that diffeomorphism invariance 
implies that all correlation functions are trivial, but this is true 
only if the abelian charges vanish. Note here the analogy with
conformal field theory \cite{FMS96}, where all interesting statistical 
systems have non-zero Virasoro central charge. The analogy is very close,
since the conformal algebra in two real dimensions is isomorphic to
(two copies of) the diffeomorphism algebra in one complex dimension.

Rudakov's theorem \cite{Rud74} states that
all proper $diff(N)$ modules are included in tensor densities; more
precisely, there is a one-to-one correspondance between $gl(N)$ and
$diff(N)$ irreps, with one exception: totally skew tensor fields, 
i.e. differential forms, contain submodules of closed differential 
forms. Hence the classical representations constructed in section 
\ref{sec:class} are highly reducible. However, this theorem does 
not apply to lowest-energy modules, since the abelian charges are
non-zero. 

Many questions remain to be answered. The most disturbing intrinsic
problems are the ambiguity in the definition of the antifields,
illustrated by the difference between (\ref{DD1}) and (\ref{DD2}),
and the problems with the limit $p\to\infty$. Another problem concerns
the decomposition into irreps; at least, every module
decomposes into its bosonic and fermionic parts.
It would also be desirable 
to make contact with the formalism of standard quantum field theory;
of course, one must then restrict to the Poincar\'e subalgbra of 
$diff(N)$.

\end{document}